\documentclass[]{spie}  %>>> use for US letter paper
\usepackage[utf8]{inputenc}
\usepackage[numbers]{natbib}
\usepackage{amsmath,amsfonts,amssymb}
\usepackage{graphicx}
\usepackage{aas_macros}
\usepackage[colorlinks=true, allcolors=blue]{hyperref}
\usepackage{gensymb}
\usepackage{xspace}
\usepackage{subcaption}
\usepackage{booktabs}
\newcommand{\expo}{\emph{EXPO}\xspace}
\newcommand{\ixpe}{\emph{IXPE}\xspace}

\title{EXPO: a quantum leap in fast, wide-band X-ray polarimetry for astrophysics}

\author[1,*]{Paolo Soffitta}
\author[2]{Sebastien Guillot}
\author[3]{Ian Hutchinson}
\author[1]{Fabio Muleri}
\author[4,56]{Mark Pearce}
\author[5]{Andrea Santangelo}
\author[6]{Daniele Spiga}
\author[7]{Ivan Agudo}
\author[8]{Gino Bruno Amata}
\author[9]{Jarosław Bakala}
% ...

\affil[1]{INAF-IAPS, Via Fosso del Cavaliere 100, 00133 Rome, Italy} 
\affil[2]{IRAP, 9 Av. du Colonel Roche BP 44346, 31028 Toulouse, Cedex 04, France}
\affil[3]{Univ. Leicester, University Road, Leicester, LE1 7RH, UK}    
\affil[4]{KTH Royal Institute of Technology, Department of Physics, 106 91 Stockholm, Sweden}
\affil[5]{Univ. Tuebingen, Geschwister-Scholl-Platz, 72074 Tübingen, Germany}
\affil[6]{INAF-OAB, via E. Bianchi 46 23807 Merate (Lc), Italy}
\affil[7]{IAA-CSIC, Glorieta de la Astronomía, s/n, 18008 Granada, Spain}
\affil[8]{Thales-Alenia Space, Strada Antica di Collegno, 253, 10146 Torino, Italy}
\affil[9]{CBK-PAN, Bartycka 18A, 00-716 Warsaw, Poland}
% ...
\affil[56]{The Oskar Klein Centre for Cosmoparticle Physics, AlbaNova University Center, 106 91 Stockholm, Sweden}
% ...

\authorinfo{Further author information: (Send correspondence to P.S.)\\P.S.: paolo.soffitta@inaf.it}

\begin{document} 
\maketitle

\begin{abstract}
The \textit{Enhanced X-ray Polarimetry Observatory} (\expo) is a mission concept proposed to ESA as an M8 candidate, with a prospective launch around 2041. Building on the scientific success of the \textit{Imaging X-ray Polarimetry Explorer} (\ixpe), \expo is designed not only to overcome its two main limitations, the narrow 2--8 keV energy band and the very slow repointing time, but also to unlock new scientific capabilities. A wide energy band and fast repointing are essential for addressing key questions in high-energy astrophysics, including the origin of the hard X-ray emission in magnetars and black-hole binaries, particle acceleration in supernova remnants and pulsar-wind nebulae, radiative transfer in highly magnetized plasmas, X-ray reflection in accretion flows and active galactic nuclei, and the prompt and afterglow emission of gamma-ray bursts and of magnetar flares.

The payload combines broadband imaging X-ray polarimetry with rapid time-domain capability. Five focusing X-ray telescopes employ gas photoelectric polarimeters based on the Timepix ASIC family with InGrid amplification, enabling true three-dimensional track reconstruction and operation over the 2--35 keV band through optimized low- and medium-energy detector configurations. The mirror modules are based on proven electroformed nickel technology with Au/C coatings and an XMM-like focal length of 7.5 m.

The polarimeters are complemented by three dedicated detection systems: (i) a coded-mask Wide Field Instrument (WFI), derived from \textit{SVOM}/ECLAIRs, providing continuous monitoring of a $\sim$2 sr field of view and autonomous transient localization; (ii) a Spectral Imaging Camera (SIC), based on stacked CMOS and CdTe detectors, providing simultaneous broadband imaging spectroscopy for accurate spectro-polarimetric decomposition; and (iii) an Instrument Control Unit (ICU), responsible for payload management, onboard WFI image reconstruction, transient identification, and autonomous spacecraft repointing. Together, these capabilities will extend X-ray polarimetry into the hard X-ray domain and open a new observational window on fast transients, time-domain astrophysics, and multi-messenger astronomy. This paper presents the concept of the mission \expo , its payload architecture, and the scientific opportunities provided by its unique combination of broadband polarimetry and rapid response.
\end{abstract}


\section{INTRODUCTION}
\label{sec:intro}  % \label{} allows reference to this section

The NASA/ASI mission \ixpe has opened the long-awaited window of soft X-ray polarimetry, demonstrating that polarization provides unique information complementary to spectroscopy, timing, and imaging for understanding the geometry and physics of compact objects and high-energy plasmas \cite{Weisskopf2022}. However, as a Small Explorer mission, \ixpe is limited by its narrow 2--8 keV energy range and slow repointing capability, preventing the study of higher-energy emission components and rapidly evolving transients. The forthcoming \textit{eXTP} mission \cite{2025extp} will improve sensitivity but will retain similar limitations in energy coverage and response time.

\textbf{EXPO} is conceived as the next-generation X-ray polarimetry observatory based on three pillars: (i) broadband polarimetry over 2--35 keV, (ii) \textit{Swift}-class autonomous repointing within tens of seconds, and (iii) simultaneous spectroscopy through a dedicated Spectral Imaging Camera (SIC). Compared to \ixpe, \expo provides a four-times broader energy band, about three-times larger collecting area, angular resolution down to $10''$, and a Wide Field Instrument (WFI) coaligned with X-ray telescopes and monitoring one sixth of the sky ($\sim2$ sr) to detect and localize transients and trigger rapid (1 degree/s) autonomous observations.

The extended energy range directly addresses several of the most important questions raised by \ixpe. It enables polarization measurements of cyclotron lines in accreting X-ray pulsars, whose unexpectedly low polarization challenges current radiation-transfer models \cite{Caiazzo2021,Caiazzo2021b}; probes the hard X-ray emission of black-hole binaries, where polarization increases with energy beyond theoretical expectations \cite{Ratheesh2024}; measures the hard tails of magnetars and low-mass X-ray binaries; investigates synchrotron emission from supernova remnants and pulsar-wind nebulae; and studies reflection components in AGN and X-ray binaries above the iron complex. Combined with improved imaging and reduced background, \expo will perform spatially resolved polarimetry of particle acceleration sites and magnetic turbulence.

Rapid repointing represents the second major breakthrough. A WFI derived from the \textit{SVOM}/ECLAIR heritage continuously monitors the sky, while onboard intelligence autonomously identifies significant variability and commands repointing. GRBs and other fast transients can be observed within $\sim50$ s, whereas slower phenomena, including state transitions of X-ray binaries, magnetar outbursts and blazar flares, are reached within minutes without hydrazine consumption. EXPO can also respond to external alerts, providing a natural interface with multi-messenger observatories.

This capability uniquely enables time-resolved spectro-polarimetry of GRB prompt emission, afterglows, and X-ray transients. Simultaneous polarization, spectroscopy, and timing will discriminate synchrotron and inverse-Compton emission mechanisms, constrain jet geometry and magnetic-field evolution, and investigate the electromagnetic counterparts of gravitational-wave events. Unlike existing or planned facilities, \expo combines broadband polarimetry, rapid response, and simultaneous spectroscopy within a single observatory.

Building on the scientific legacy of \ixpe, \expo is aimed at transforming X-ray polarimetry from a discovery mission into a mature observatory, opening entirely new research domains and placing Europe at the forefront of high-energy astrophysics in the post-\textit{NewAthena} era.

\section{What EXPO can do that IXPE cannot}
\begin{figure}[ht!]
    \centering
    \includegraphics[width=1.0\linewidth]{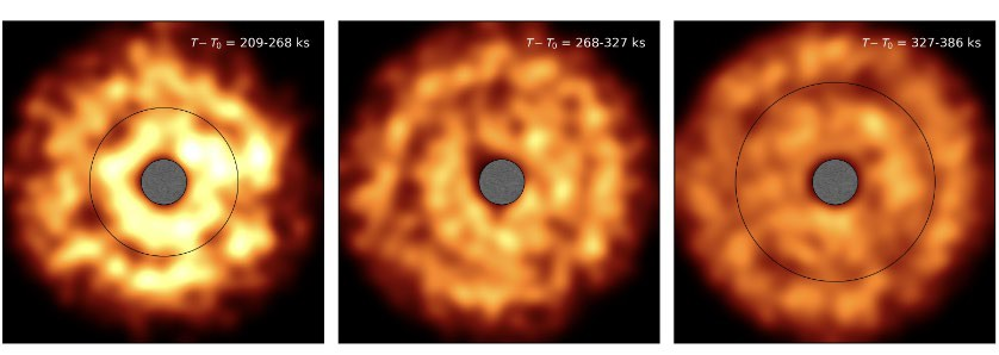}
    \caption{The dust scattering echo revealed by \ixpe which allowed the observation of the BOAT prompt emission and the determination of the upper limit in polarization \cite{Negro2023}}
    \label{fig:BOAT}
\end{figure}

One of the most compelling examples illustrating the scientific need for an observatory such as \expo is GRB 221009A, detected by \textit{Swift} on 9 October 2022 and soon dubbed the \emph{Brightest Of All Time} (BOAT, see Figure \ref{fig:BOAT}). Due to a fortunate coincidence between the issuance of the ATel and the weekly mission replanning cycle, \ixpe was able to repoint toward the burst only three days after the trigger. Another fortunate circumstance was the location of the burst, which produced dust-scattering echo rings in the Galactic Plane, detected by both \textit{Swift} and \ixpe. These expanding rings made it possible to spatially separate the scattered prompt emission from the direct afterglow, allowing an independent search for polarization in both components.

Despite this unique opportunity, the limited collecting area of \ixpe , and the inadequate repointing time resulted only in weak constraints on polarization. At the 99\% confidence level, the upper limit on the degree of polarization of the afterglow was 13.8\%, while the corresponding upper limits for prompt emission were approximately 55\% and 82\%, depending on the adopted background model. These limits are well above the polarization levels predicted by most theoretical models (potentially up to 60\% \cite{2003Granot}) and therefore cannot discriminate among competing emission mechanisms.

With its substantially larger effective area, wider energy coverage, and rapid autonomous repointing capability, \expo will transform GRBs events into precision measurements, providing decisive constraints on the geometry, magnetic-field structure, and radiation processes operating during both the prompt and afterglow phases of GRBs.

\begin{figure}[ht!]
    \centering
    \includegraphics[width=0.5\linewidth]{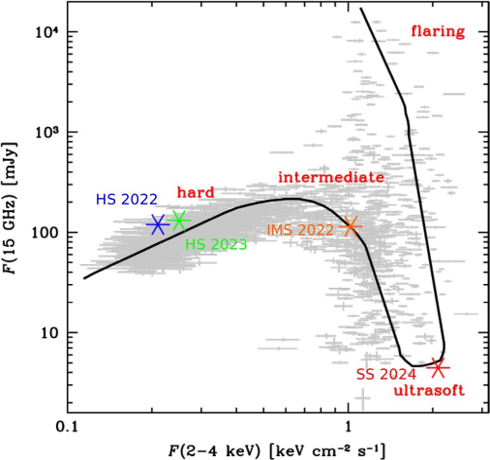}
    \caption{State-dependent evolution of Cyg~X-3 \cite{Mikusincova2026}. The hard and hard-intermediate states may evolve into a radio-quenched (high-soft) state, which is believed to precede major radio flares associated with the ejection of relativistic plasmoids during jet formation.}
    \label{fig:Cyg X-3}
\end{figure}

Another key science objective for \expo, beyond the capabilities of \ixpe, is the investigation of jet formation in microquasars through rapid repointing triggered by major radio flares. Such flares are believed to mark the ejection of relativistic plasmoids along the system's rotation axis and provide a unique opportunity to probe the geometry and physical conditions of the innermost accretion flow during jet launching. A recent example is provided by Cyg~X-3. Although \ixpe observed this source several times \cite{Veledina2023,Veledina2024, Veledina2024b, Mikusincova2026} throughout its radio evolution (see Figure \ref{fig:Cyg X-3}), none of the observations could be made during or immediately following a major radio flare, when the jet formation process is expected to occur.

Understanding how relativistic jets are launched in microquasars has implications that extend well beyond Galactic X-ray binaries. These systems constitute nearby laboratories for the study of the accretion--ejection coupling operating in AGNs and may also provide valuable insights into the past activity of the supermassive black hole at the Galactic center. By combining rapid autonomous repointing with broadband X-ray polarimetry, \expo will uniquely probe the geometry of the accretion funnel and the physical conditions that trigger jet launching during these short-lived impulsive events. 

\begin{figure}[ht!]
    \centering
    \includegraphics[width=0.5\linewidth]{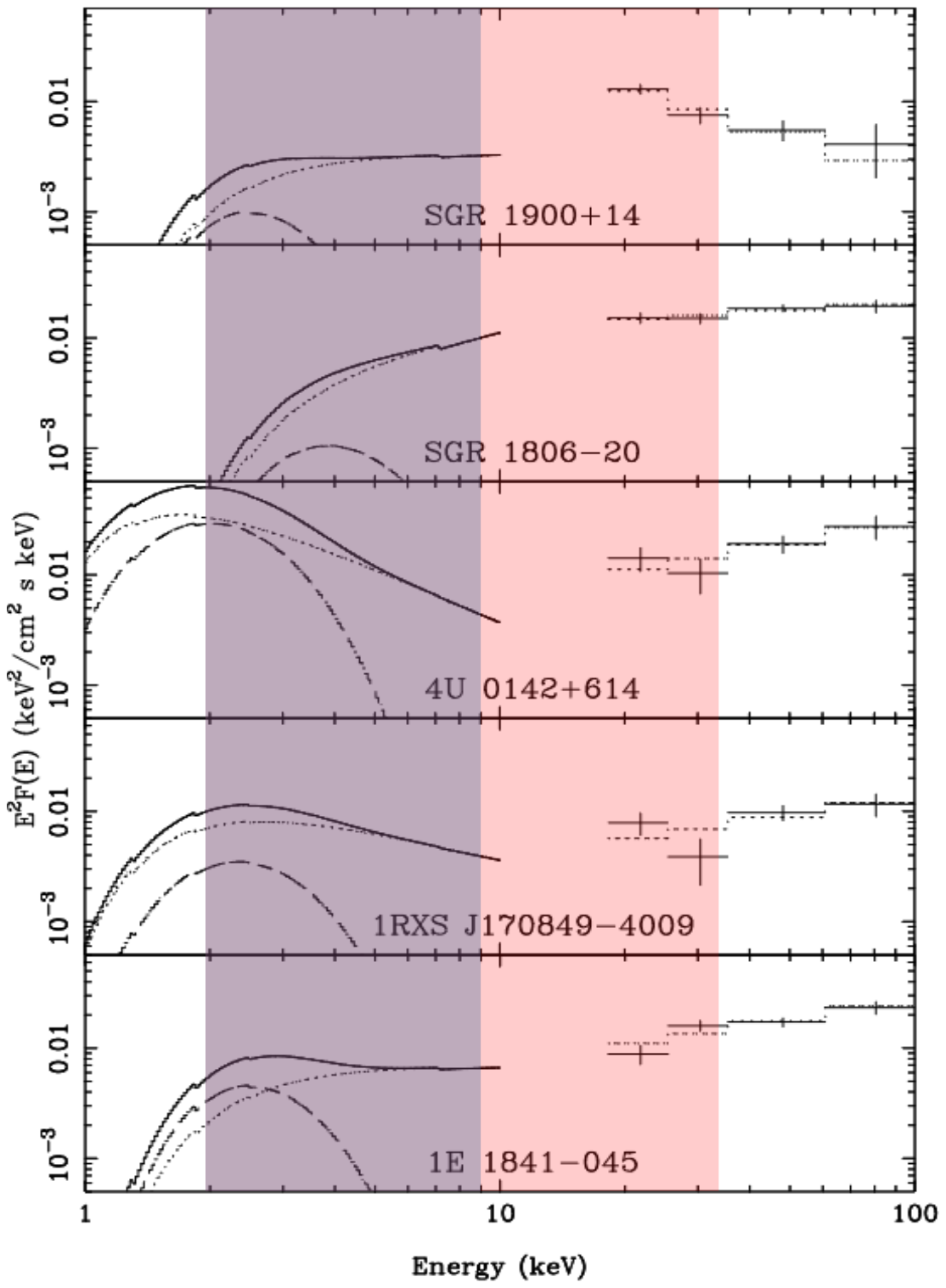}
    \caption{A sample of magnetar spectra (adapted from \cite{Gotz2006}) illustrating their hard X-ray tails, with the \ixpe (purple) and \expo (purple and pink) energy bands overlaid. The spectral connection between the soft and hard X-ray components differs markedly among sources, emphasizing the need for the broadband polarimetric capabilities of \expo to establish the origin of the hard X-ray emission tail.}
    \label{fig:Magnetars}
\end{figure}

Another compelling science case for \expo is the determination of the physical origin of the hard X-ray tail observed in magnetars (see Figure \ref{fig:Magnetars} adapted from \cite{Gotz2006}), which extends from $\sim10$ keV to several hundred keV. This emission is almost entirely outside the 2--8 keV \ixpe energy range, while it falls naturally within the bandwidth capabilities of \expo.
The origin of this hard component remains one of the major unsolved problems in neutron-star astrophysics. Competing models attribute it either to thermal bremsstrahlung produced by relativistic electrons in the neutron-star surface layers, synchrotron emission from electron--positron pairs accelerated in the strongly twisted magnetosphere and finally to Compton resonant up-scattering from these relativistic charges. These scenarios predict markedly different polarization signatures, making broadband X-ray polarimetry the most powerful diagnostic to distinguish them.

\begin{figure}
    \centering
    \includegraphics[width=0.9\linewidth]{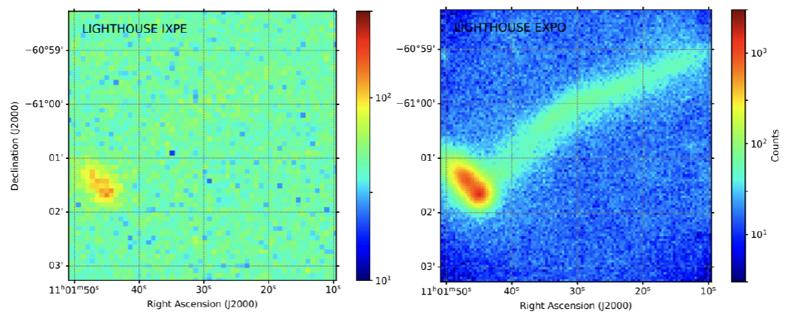}
    \caption{The Lighthouse PWN as observed by \ixpe (left) and simulated for \expo (right) with the same 1~Ms exposure. \expo will perform spatially resolved polarimetry of the filament, mapping the magnetic-field orientation and extending these studies to many PWNe beyond the reach of \ixpe .}
    \label{fig:lighthouse}
\end{figure}

An illustrative example, Figure~\ref{fig:lighthouse} compares simulated \ixpe and \expo observations of the non-thermal filament attached to the Pulsar Wind Nebula (PWN) known as the \emph{Lighthouse}. This remarkable system \cite{Halpern2014} exhibits a radio and X-ray trail extending opposite to the pulsar proper motion, while a bow shock is formed ahead of the pulsar. An even more intriguing feature is a bright X-ray filament, with no radio counterpart, oriented at a large angle with respect to the trail and accompanied by a fainter counter-filament, as revealed by \textit{Chandra} observations \cite{Dinsmore2024}. The prevailing interpretation is that these structures trace relativistic leptons escaping through the bow shock into the surrounding interstellar medium. 

Figure~\ref{fig:lighthouse} compares an \ixpe image with the simulated \expo image obtained for a similar observing time, highlighting the dramatic improvement in spatial detail. With \ixpe, polarization could only be measured by integrating over the entire filament, yielding a polarization angle consistent, at the 99\% confidence level, with a magnetic field aligned with its elongation. Thanks to its superior angular resolution and larger effective area, \expo will resolve the polarization properties along the filament, enabling the magnetic-field orientation to be mapped as a function of position and providing a direct test of the hypothesis that the structure is produced by relativistic particles escaping from the bow shock into the interstellar medium. Similar studies will become feasible for a significant sample of PWNe, extending polarimetric investigations to higher energies and to sources beyond the reach of \ixpe .

\begin{figure}[ht!]
    \centering
    \includegraphics[width=0.45\linewidth]{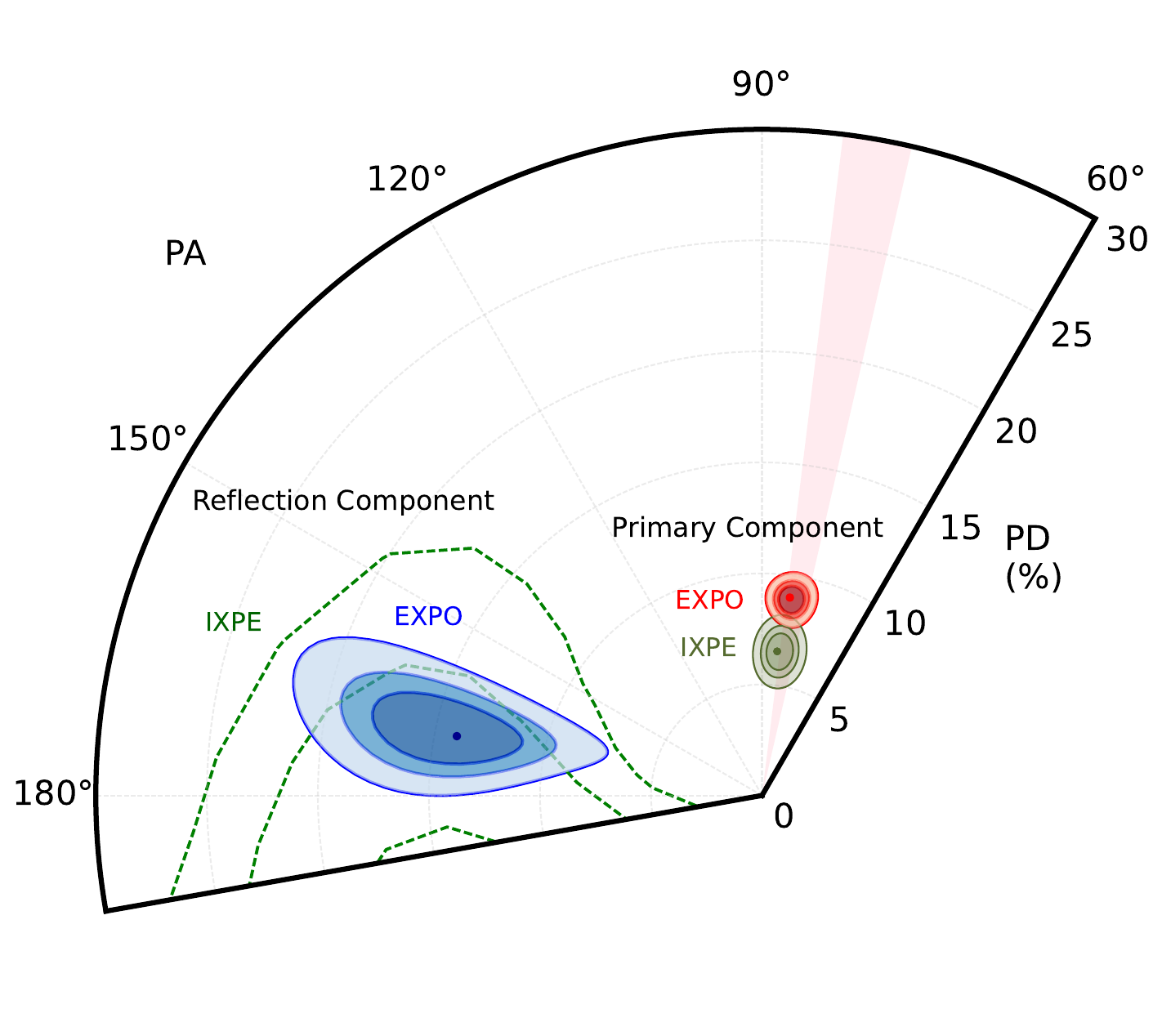}
    \caption{Comparison of the \ixpe observation and the simulated \expo observation of NGC~4151. 
    Contours are at 68, 90, and 99\% c.ls. \ixpe results are shown in green, \expo simulations (assuming a PD of the primary/reflection component of 7\%/10\%, the latter a rather conservative assumption) are shown in red for the primary continuum and blue for the reflection component. The pink bands in the polar plots indicate the direction of the radio emission.
    The wider energy band of \expo enables a detailed measurement of the polarization of the disk reflection component, beyond the reach of \ixpe.}
    \label{fig:NGC4151}
\end{figure}

An illustrative example of the scientific gain enabled by a wide-band X-ray polarimetry mission such as \expo is provided by the simulated observation of NGC~4151, the archetypal Seyfert~1 galaxy for which \ixpe measured a significant X-ray polarization degree \cite{Gianolli2024}. The measured polarization angle is aligned with the faint radio jet, supporting a coronal geometry elongated parallel to the accretion disk. A major advance offered by \expo arises from its broad energy coverage, which allows the reflected spectral component to be explored with high precision. 

As shown in Figure~\ref{fig:NGC4151}, the reflection component, which remains largely unconstrained by \ixpe over the same observing time because of its limited energy range, can instead be accurately characterized by \expo. This capability will provide stringent constraints on the geometry of the X-ray corona, the properties of the reflecting disk, and the effects of strong gravity in the immediate vicinity of the supermassive black hole.

%As can be seen in the plot, the component of reflection in the spectrum can be efficiently explored by \expo thanks to its wider energy band being instead unconstrained by \ixpe in the same observing period.

\section{Main features of EXPO}
\expo is expected to exceed \ixpe in several technical and scientific aspects (see Table \ref{tab:EXPOoverIXPE}). This is possible thanks to a suite of instruments and spacecraft features that ensure that science requirements are truly met. The main payload elements already have an extensive heritage, while all are expected to reach a technological readiness level (TRL) of 5 at the end of phase A.

\begin{table}[ht!]
\caption{What \expo improves over \ixpe.}
\label{tab:EXPOoverIXPE}
\begin{center}       
\begin{tabular}{|l|l|} %% this creates two columns
%% |l|l| to left justify each column entry
%% |c|c| to center each column entry
%% use of \rule[]{}{} below opens up each row
\hline
\rule[-1ex]{0pt}{3.5ex}  Wider energy band &  4 times  \\
\hline
\rule[-1ex]{0pt}{3.5ex}  Increased effective area &  3 times in the overlapping band \\
\hline
\rule[-1ex]{0pt}{3.5ex}  Better angular resolution &  1/3 of \ixpe (HEW 10'' vs 30'')  \\
\hline
\rule[-1ex]{0pt}{3.5ex}  Lower background rate &  20 times lower\\
\hline
\rule[-1ex]{0pt}{3.5ex}  Better response to Target of Opportunities &  8000 times faster  \\
\hline
\rule[-1ex]{0pt}{3.5ex}  Field of view for autonomous discovery of transients &  in 1/6 of the sky  \\
\hline
\rule[-1ex]{0pt}{3.5ex}  Auxiliary Spectral capability &  with a CMOS $\&$ CZT energy resolution \\
\hline
\end{tabular}
\end{center}
\end{table}

\begin{figure}
\centering
\includegraphics[width=0.64\textwidth]{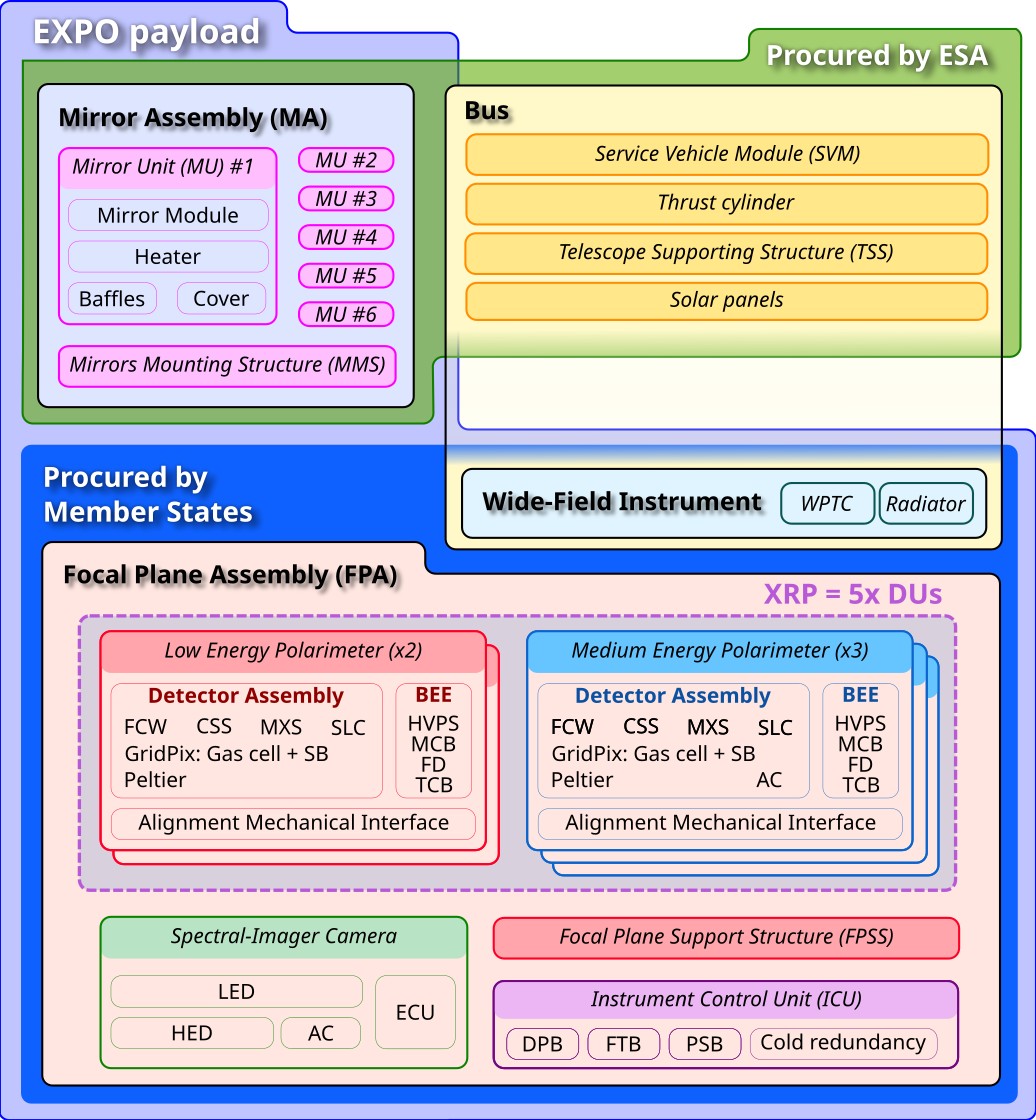}
\caption{Functional diagram of the \expo payload, showing the distribution of the payload elements between the Service Module (SVM) and the Focal Plane Assembly (FPA). The subdivision between ESA procured and Member State procured items is also shown.}
\label{fig:EXPO_blockdiagram}
\end{figure}

The \expo payload, shown in Figure~\ref{fig:EXPO_blockdiagram}, is distributed between the Service Module and the Focal Plane Support Structure (FPSS). Its design is based on two complementary pillars: broadband X-ray polarimetry and rapid-response time-domain astronomy which we describe below.

\textbf{Extending the energy band:}
Broadband X-ray polarimetry requires gas detectors optimized for different energies, since the gas composition, pressure, and absorption depth determine the detector performance and characteristics. \expo therefore employs two complementary Gas Pixel Polarimeters: two \textbf{Low Energy Polarimeters (LEPs)}, derived from the successful \ixpe design, and three \textbf{Medium Energy Polarimeters (MEPs)} optimized for hard X-rays. The six telescopes have a common 7.5 m focal length, comparable to that of \textit{XMM-Newton}, providing the shallow grazing angles required to cover the full energy range.

The LEPs employ 1 atm mixtures based on low-$Z$ gasses (like Ne), while the MEPs use a 30 mm absorption gap filled with an Ar-based mixture at 3 atm to maximize efficiency above $\sim$8 keV. Both detector types are based on an InGrid amplification stage coupled to a Timepix3 ASIC. % (Fig.~\ref{subfigure:InGridPicture}).

A sixth telescope hosts the \textbf{Spectral Imaging Camera (SIC)}, which combines a silicon detector and a stacked CZT detector to provide simultaneous spectroscopy over a broader energy range with significantly improved spectral resolution. The high-quality spectra acquired simultaneously with the polarimeters enable robust decomposition of the different spectral components, substantially improving polarization measurements.

\textbf{A new era of time-domain polarimetry:}
X-ray sources exhibit strong variability in flux, spectrum, and polarization, making rapid response essential. While \ixpe had limited observing flexibility and a small number of Targets of Opportunity, \expo is designed for rapid autonomous operations. A dedicated \textbf{Wide Field Instrument (WFI)} continuously monitors nearly two steradians of the sky, detecting transients on timescales of seconds and triggering autonomous spacecraft repointing within about 50 s. This capability opens the largely unexplored field of polarimetry of fast transients, including Gamma-Ray Bursts.

Rapid repointing can also be initiated by external alerts, making \expo a key facility for multi-messenger astronomy. The SIC provides accurate source localization within the focal plane, while the payload is coordinated by the \textbf{Instrument Control Unit (ICU)} and mechanically integrated with the spacecraft through the \textbf{Focal Plane Support Structure (FPSS)}.

\section{The Mirrors and the Wide Field Instrument in the Service Module}

This section describes the payload elements accommodated within the Service Module (SVM).

\subsection{X-ray Mirrors}
\begin{figure}[ht!]
    \centering
    \begin{tabular}{cc}
    \includegraphics[width=0.40\textwidth]{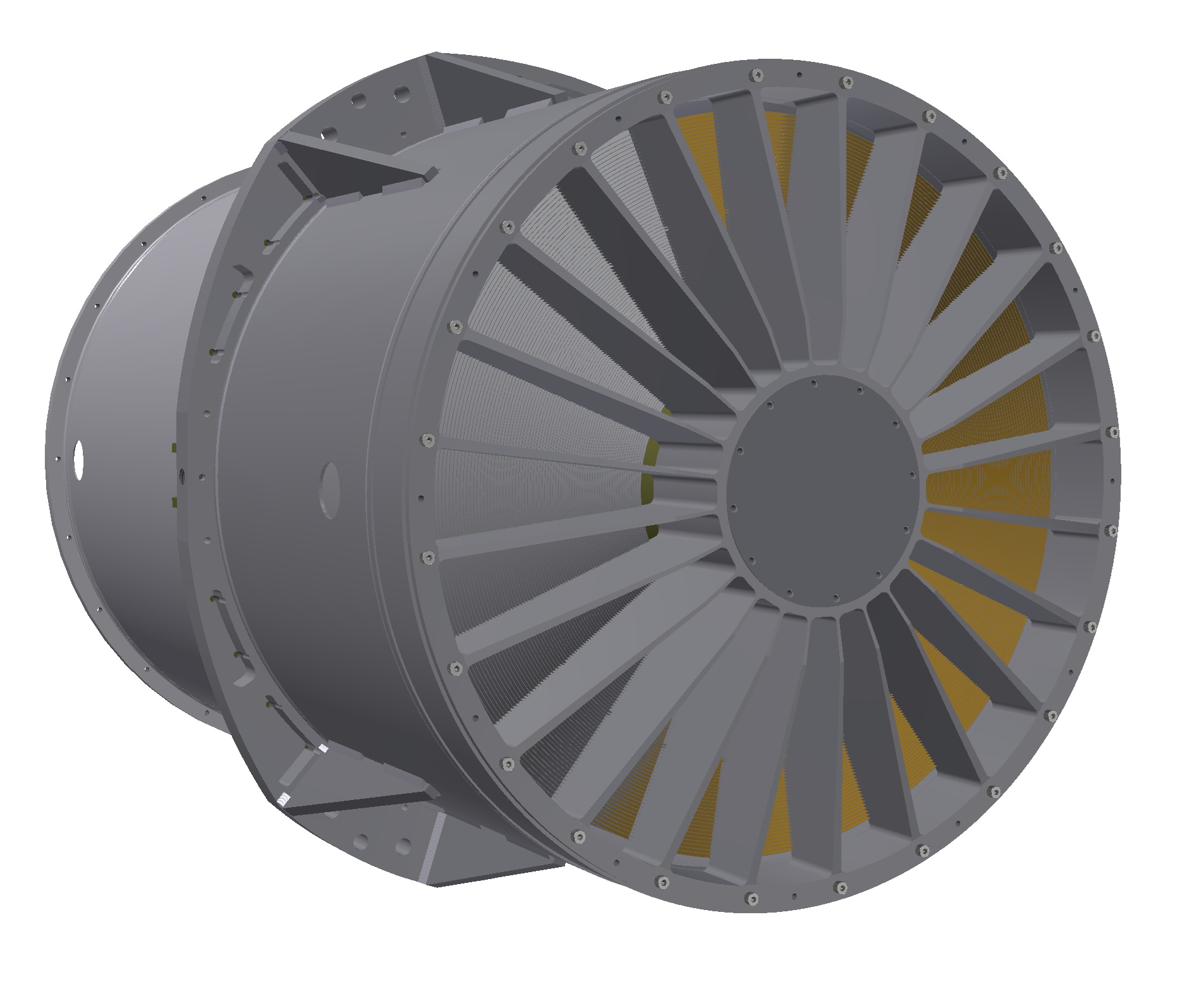}
    &
    \includegraphics[width=0.50\textwidth]{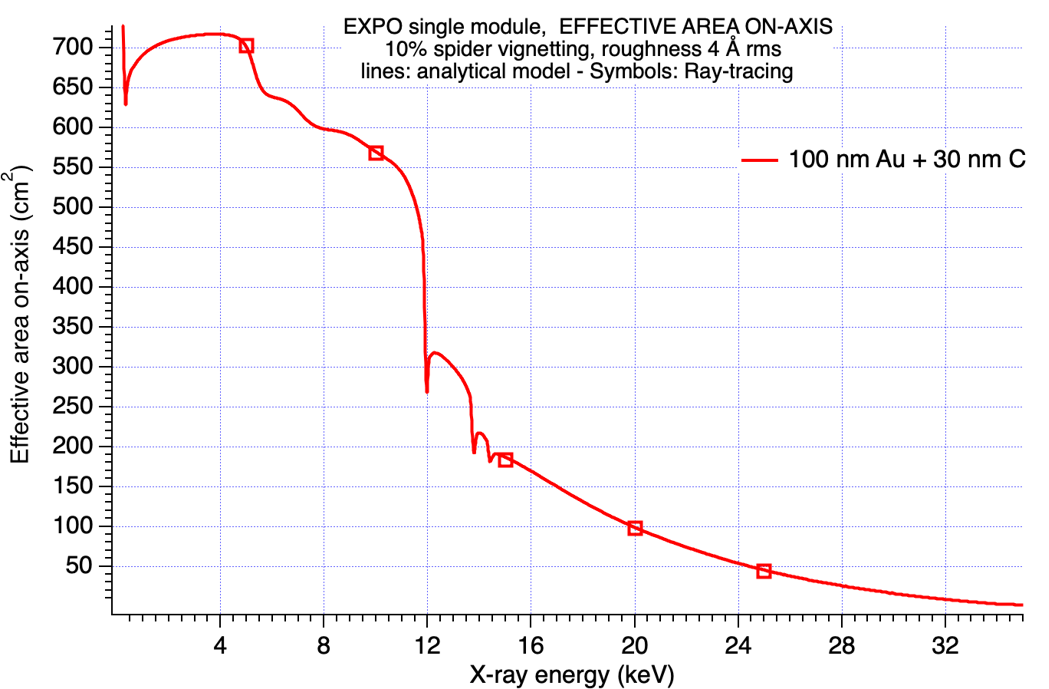}
    \end{tabular}
    \caption{\footnotesize{Left: \expo mirror unit design. Right: mirror unit on-axis effective area.}}
    \label{fig:Module&Area&EffArea}
\end{figure}

The six X-ray Mirror Units (MUs) are mounted on an Optical Bench (OB) located within the central aperture of the Service Module (SVM).
The \expo telescopes are based on nested Wolter-I grazing-incidence mirrors, whose main parameters are described in Table \ref{tab:mirror_parameters}. They are manufactured by nickel electroforming from super-polished mandrels, a technology successfully employed from \textit{BeppoSAX} and \textit{XMM-Newton} to \textit{Swift}, \textit{SRG/eROSITA}, and \ixpe. The mirror modules (see Figure \ref{fig:Module&Area&EffArea}) will be produced by Media Lario using their well-established expertise in mandrel fabrication, shell electroforming, and optical integration. Each module comprises 80 nested shells coated with Au and a thin C overcoating. Gold provides high reflectivity up to $\sim$35 keV, while the carbon layer fills the Au M-edge reflectivity losses at low energies. The carbon coating, based on dopamine deposition, is currently under development at INAF--Brera with ASI support \cite{Cotroneo2022,Khropost2025}.

\begin{table}[htbp]
\caption{Main characteristics of a single \expo mirror module.}
\label{tab:mirror_parameters}
\centering
\begin{tabular}{ll}
\hline
Parameter & Value \\
\hline
Optical design & Wolter-I grazing incidence \\
Manufacturing & Ni electroformed replica \\
Number of shells & 80 \\
Shell diameter & 400--138.5 mm \\
Shell length & 600 mm \\
Focal length & 7500 mm \\
Incidence angle & $0.38^\circ$--$0.13^\circ$ \\
Reflective coating & Au + C overcoating \\
Energy range & 2--35 keV \\
Field of view & 13 arcmin (50\% vignetting at 10 keV) \\
Angular resolution & 10 arcsec HEW @ 1 keV \\
 & 14 arcsec HEW @ 10 keV \\
Stray light & $<1\%$ on-axis effective area \\
Mirror mass & 89.1 kg \\
Module mass & 121.8 kg \\
\hline
\end{tabular}
\end{table}

The mirror design provides a 13 arcmin field of view (50\% vignetting at 10 keV), stray-light below 1\% of the on-axis effective area (which is shown in Figure \ref{fig:Module&Area&EffArea}), and an angular resolution of 10 arcsec HEW at 1 keV, increasing only to $\sim$14 arcsec at 10 keV thanks to the excellent mandrel surface quality. Shell integration follows the proven \textit{SRG/eROSITA} gravity-compensation procedure using suspended shells and UV-assisted alignment to ensure accurate co-focality.

%

%\begin{figure}
%    \centering
%    \includegraphics[width=0.5\linewidth]{EffArea.png}
%    \caption{Enter Caption}
%    \label{fig:EffArea}
%\end{figure}
%\begin{figure}
%    \centering
%    \includegraphics[width=0.5\linewidth]{EXPO1.jpg}
%    \caption{Enter Caption}
%    \label{fig:placeholder}
%\end{figure}

\subsection{Wide Field Instrument (WFI)}

The \textbf{Wide Field Instrument (WFI)}, provided by IRAP (France), is a coded-mask imager derived from the successful \textit{SVOM}/ECLAIRs instrument \cite{Atteia2022}. Its primary role is the detection and localization of high-energy transients, followed by the autonomous repointing of \expo for rapid observations with the narrow-field instruments. The WFI operates in the 4--150 keV band using a CdTe detection plane and a pseudo-random coded mask covering a field of view of about 2 sr. Continuous onboard processing of count-rate and imaging triggers enables real-time transient identification and makes the WFI the cornerstone of the time-domain capability of the mission. The main instrument characteristics are summarized in Table~\ref{tab:WFIPerformances}.

\begin{table}{r}
% \begin{table}[h]
\centering
\caption{Characteristics of the WFI.} 
\label{tab:WFIPerformances} 
\begin{tabular}{ll}
\hline
\textbf{Parameter} & \textbf{WFI Characteristic} \\
\hline
Energy range & 4.0--150 keV \\
Angular resolution & 1.5$^{\circ}$ \\
Source localization & $<$ 12'' \\ % Note: changed to \arcsec if you loaded AAS macros
Field of view & $89^{\circ}\times 89^{\circ}$ (2.05 sr) \\
Energy resolution (FWHM) & $\sim$ 1.2 keV (@ 60 keV) \\
Sensitivity (over 1 s in 5--50~keV) & 2.5$\times10^{-8}$ erg cm$^{-2}$s$^{-1}$\\
Number of GRBs per year & $\sim$65 \\
Detection plane area & $\sim$ 950 cm$^{2}$ \\
Dead time & $<$10\% for $10^{5}$ c/s\\
\hline
\end{tabular}
\end{table}

The coded mask consists of a $54\times54$ cm$^2$ Ta sheet reinforced by titanium, with a 40\% open fraction and a mask-to-detector distance of 46 cm, providing an $89^\circ\times89^\circ$ field of view. A Cu--Pb passive shield suppresses photons below $\sim$50 keV entering from outside the field of view.

The detector plane comprises 6400 CdTe pixels arranged in 200 modules, each read out by the low-power IDeF-X ASIC. Photons above the $\sim$4 keV threshold are digitized to 10-bit precision and 20 $\mu$s timing accuracy. Unlike ECLAIRs, the real-time transient search will be performed by the \expo Instrument Control Unit (ICU), while a dedicated WFI Power and Thermal Control (WPTC) unit, located on the Service Module, provides detector bias and thermal regulation.

Thanks to its direct heritage from \textit{SVOM}/ECLAIRs, the WFI has a high projected TRL and presents no major technological risks. The only significant interface is the high-speed data link to the ICU, carrying an average data volume of about 15 Gbit day$^{-1}$. The WFI requires an unobstructed sky view and passive radiator cooling but no precise co-alignment with the focusing telescopes, since its role is to detect transients and trigger spacecraft repointing.

\subsection{The service module and the location of payload items}
\begin{figure}[ht!]
    \centering
    \includegraphics[width=0.5\linewidth]{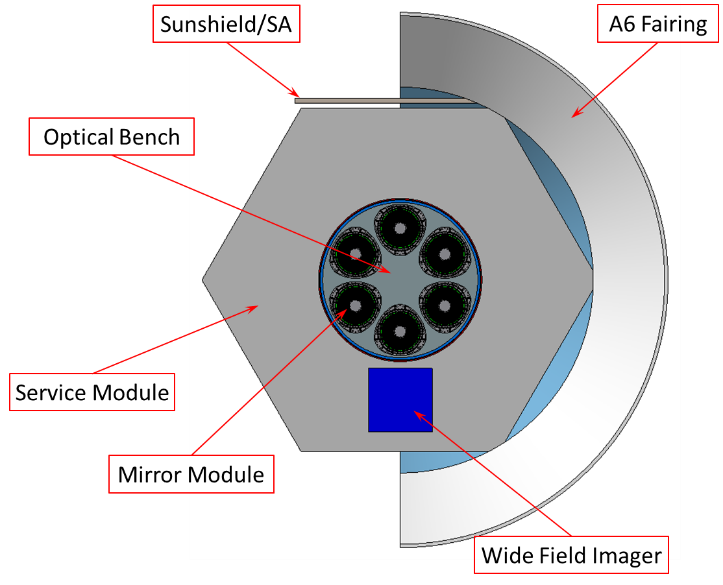}
    \caption{The Service Module with the accomodation of the WFI and the 6 mirrors}
    \label{fig:FrontPanel}
\end{figure}

Thales Alenia Space Torino performed a preliminary accommodation study of the six mirror units (MUs) and the Wide Field Instrument (WFI) within the proposed Service Module (see Figure \ref{fig:FrontPanel}). In the initial configuration, the WFI was located at the centre of the circular arrangement of the mirror modules. The baseline design was subsequently revised by relocating the WFI to the edge of the Service Module. This solution offers several advantages: (i) a shorter and simpler interface to the spacecraft services, (ii) a more efficient thermal design through direct coupling to dedicated radiators, and (iii) increased accommodation margin, simplifying the overall mechanical integration.

The six Mirror Units are accommodated within the available aperture of the launcher interface adapter with comfortable mechanical margins, taking into account the mirror housings, mounting flanges, and bolted interfaces. The resulting configuration provides a robust and manufacturable payload layout while preserving sufficient clearance for integration, alignment, and spacecraft interfaces. 

A telescope tube connects the optical bench within the spacecraft, where the optics are fixed and aligned, to the FPSS providing the necessary 7.5 meter focal length.

The Service Module (SVM) proposed by Thales Alenia Space Torino, together with the accommodation of the payload elements, is shown in Figure~\ref{fig:FrontPanel}. The complete spacecraft is housed within the Ariane 62 fairing with comfortable mechanical margins.

\section{The Focal Plane Support Structure and the location of the Payload Items}

The detectors located in the focal plane are the X-ray Polarimeters (XRP) and the Spectral-Imaging Camera. The XRP are composed of two Low Energy Polarimeters (LEP) and three Medium Energy Polarimeters (MEP). The sixth detector is the Spectral-Imaging Camera (SIC) composed of a stacked system with a CMOS above a pixelated CdTe/CZT detector. They are all centered with respect to the X-ray mirror focal axis and share the common focal plane. The last but not least payload/item of the focal plane is the Instrument Control Unit which serves to provide power to XRP and SIC, to manage the payload modes and event timing and manage the WFI autonomous repointing.

\subsection{Photoelectric Polarimeters}

The X-ray polarimeters (XRPs) are the core of the \expo payload. They exploit the photoelectric effect, as successfully demonstrated by \ixpe, but introduce major technological advances enabled by two decades of microelectronics development. When an X-ray photon is absorbed in the gas, the emitted photoelectron leaves an ionization track whose initial direction follows the polarization of the photon through a distribution $\cos^2$ \cite{Heitler1954}. By reconstructing the three-dimensional track, the interaction point, absorption energy, and photoelectron emission direction are simultaneously measured, providing imaging, spectroscopy, and polarimetry.
Thanks to the 10--14 arcsec angular resolution of the optics, detector spatial resolution and inclined penetration contribute negligibly to the overall imaging performance.

Both the Low Energy Polarimeter (LEP) and the Medium Energy Polarimeter (MEP) are based on the GridPix concept, combining the InGrid micromesh amplification stage with the Timepix3 ASIC \cite{vanderGraaf2007,Kaminski2012,Poikela2014,Krieger2013}. Compared with the \ixpe detector, this architecture provides true 3-D track reconstruction, negligible dead time, digital pixel readout, and virtually diffusion-free charge multiplication owing to the precise alignment between the InGrid holes and the ASIC pixels. The main characteristics of LEP and MEP are reported in Table \ref{tab:LEPMEPOerformances}.

To cover the 2--35 keV band, the two detectors adopt different gas cells. The LEP uses a 10 mm cell filled with Ne/DME (80/20) at 1 bar, optimized for soft X-rays, whereas the MEP employs a 30 mm Ar/DME (60/40) cell at 3 bar for high-energy photons. Five Detector Units (two LEPs and three MEPs) are foreseen. Each comprises a Detector Assembly with the GridPix sensor, Filter and Calibration Wheel, Modulated X-ray Source, thermal control and, for the MEP only, a GAGG anti-coincidence system, together with a dedicated Back-End Electronics.

\begin{table}[!htbp]
\centering
\caption{Main performance parameters of the LEP and MEP detectors. $^{\ddagger}$ From Monte-Carlo using the \ixpe algorithm on the 2-D projection of the track. $^{*}$lab measurements.}
\centering
\footnotesize
\begin{tabular}{lll}
\hline
\textbf{Parameter} & \textbf{LEP (1 detector)} & \textbf{MEP (1 detector)} \\
\hline
Energy band (polarimetric) & 2--10 keV & 6--35 keV \\
Energy band (total) & 1.0--12 keV & 2.0--35 keV \\
Spatial resolution & 100 $\mu$m & 150 $\mu$m \\
Angular resolution & 10$''$ & 15$''$ \\
Energy resolution (FWHM) & 1 keV (@ 5.9 keV) & 1 keV (@ 8.7 keV)$^{*}$ \\
Dead time & Negligible up to 40 Mhits cm$^{-2}$s$^{-1}$ & Negligible up to 40 Mhits cm$^{-2}$s$^{-1}$ \\
Drift time (from window) & 1 $\mu$s & 2 $\mu$s \\
Quantum efficiency & 22\% (@ 2.3 keV) & 58\% (@ 8.7 keV) \\
Modulation factor$^{\ddagger}$ & 34\% (@ 3 keV) & 49\% (@ 17.4 keV) \\
Internal Background rate (cts~$s^{-1}$~cm$^{-2}$~keV$^{-1}$)$^{+}$  &  1.7 $\times$ $10^{-4}$ & 5.0 $\times$ $10^{-4}$\\
\hline
\end{tabular}
\label{tab:LEPMEPOerformances}
\end{table}

\begin{figure}[ht!]
    \centering
    \includegraphics[width=0.5\linewidth]{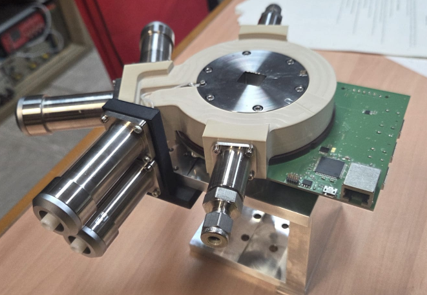}
    \caption{Prototype GridPix X-ray polarimeters undergoing testing at INAF-IAPS.}
    \label{fig:iaps-proto}
\end{figure}

The prototype GridPix detectors developed at INAF-IAPS (see Fig. \ref{fig:iaps-proto}) have already demonstrated \cite{Ratheesh2024b, Muleri2026, Manikantan2026} 3-D track reconstruction, good modulation measurements, and dead-time-free operation at count rates corresponding to several tens of Crab. The expected detector performance is summarized in Table~\ref{tab:LEPMEPOerformances}. The 3-D event topology, combined with the sparse readout and the MEP anti-coincidence system, is expected to suppress the particle background \cite{baracchini2026} by more than 90\%, resulting in the background levels reported in Table~\ref{tab:LEPMEPOerformances}.

\subsection{Spectral Imaging Camera (SIC)}
The \textbf{Spectral Imaging Camera (SIC)}, provided by the University of Leicester, the University of T\"ubingen and their partners, complements the polarimeters by providing simultaneous imaging spectroscopy over the 0.5--35 keV band. Its primary role is to constrain the spectral components of the observed sources, enabling a more efficient interpretation of the polarization measurements. The main performance parameters are listed in Table~\ref{tab:SICPerformances}.
\begin{table}[ht!]
\centering
\footnotesize
\caption{SIC characteristic.}
\label{tab:SICPerformances}
\begin{tabular}{ll}
\hline
\textbf{Parameter} & \textbf{SIC Characteristic} \\
\hline
Energy range & 0.5--35 keV \\
HEW & 10" ($<$ 10 keV) \\
 & 15" ($>$ 10 keV)\\
Mirror resolution & 10--15" \\
Mirror area &  $700$ cm$^{2}$ (@ 5 keV) \\
Field of view & $\sim$ 16' \\
Energy resolution (FWHM) & LED: $\sim 180$ eV (@ 5.9 keV)\\
 & HED: $\sim$1 keV (@ 17.5 keV)  \\
Time resolution & LED (window mode): $<5$ ms\\
 & HED: $<5$ $\mu$s \\
Detection plane area & $\sim$ 4 $\times$ 4 cm$^{2}$ \\
\hline
\end{tabular}
\end{table}

The SIC employs a stacking detector architecture consisting of a pixellated CMOS \textbf{Low Energy Detector (LED)} operating between 0.5 and 10 keV, and a pixellated CZT \textbf{High Energy Detector (HED).} that extends the response to 35 keV. Low energy photons are absorbed in the LED, while higher-energy photons are detected by the HED. An active GAGG anti-coincidence system surrounding the HED reduces the instrumental background. Both detectors share common Front-End Electronics providing a single interface to the Instrument Control Unit (ICU). The performance of the SIC is described in Table \ref{tab:SICPerformances}. 

\subsection{The Instrument Control Unit}

The \textbf{Instrument Control Unit (ICU)}, developed by the University of T\"ubingen, is the central computer of the \expo payload. It provides the interface between the Spacecraft Service Module (SVM) and all scientific instruments, performing power distribution, instrument control, science data processing, telemetry and telecommand handling, and health monitoring. The ICU also implements the mission autonomous functions, including transient identification and fast spacecraft repointing requests.

Science and housekeeping data from all instruments are collected, synchronized, monitored, and processed by the ICU before transmission to the spacecraft mass memory. In addition to standard telemetry products, the ICU executes dedicated algorithms, including image deconvolution for the WFI and transient detection, enabling the rapid-response capability of \expo.

The ICU architecture is based on three cold-redundant board types: the \textbf{Data Processing Board (DPB)}, hosting the central processor and mass memory; the \textbf{Filter Wheel and Thermal Control Board (FTB)}, controlling the filter wheels and thermal subsystems; and the \textbf{Power Supply Board (PSB)}, responsible for regulated power generation and distribution. Internal communication relies on a CAN bus, while SpaceWire links connect the ICU to the instruments and the spacecraft.

The ICU is mounted approximately at the centre of the rear side of the FPSS. To support the Assembly, Integration and Verification (AIV) flow, it is integrated before the XRPs and SIC, allowing early functional testing of the payload electrical architecture. 

\subsection{The Focal Plane Assembly}
\begin{figure}[htbp!]
\caption{Focal plane elements.}
\centering
\begin{subfigure}[t]{0.15\textwidth}
\centering
\includegraphics[width=\linewidth]{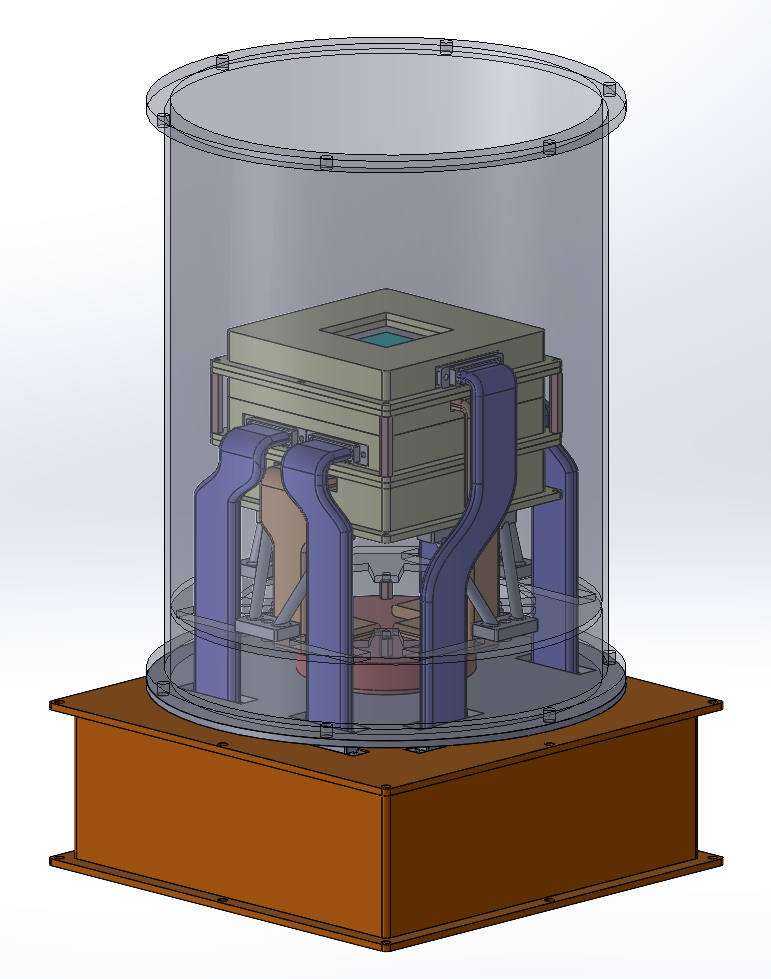}
\caption{The SIC design with detector stacking system, the thermal items and the Back-End Electronics on the bottom.}
\label{subfigure:SIC}
\end{subfigure}
\hfill
\begin{subfigure}[t]{0.45\textwidth}
\centering
\includegraphics[width=\linewidth]{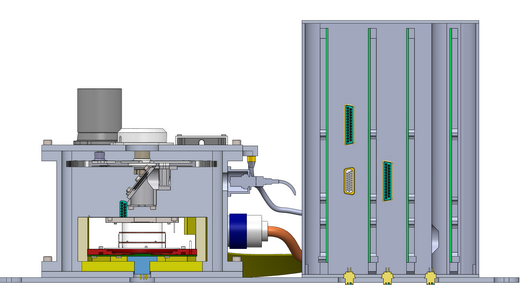}
\caption{The MEP detector with its Field Forming Rings, the GAGG anti-coincidence system and the MXS based calibration system (on the left) and its electronics (on the right).}
\label{subfigure:MEP}
\end{subfigure}
\hfill
\begin{subfigure}[t]{0.35\textwidth}
\centering
\includegraphics[width=\linewidth]{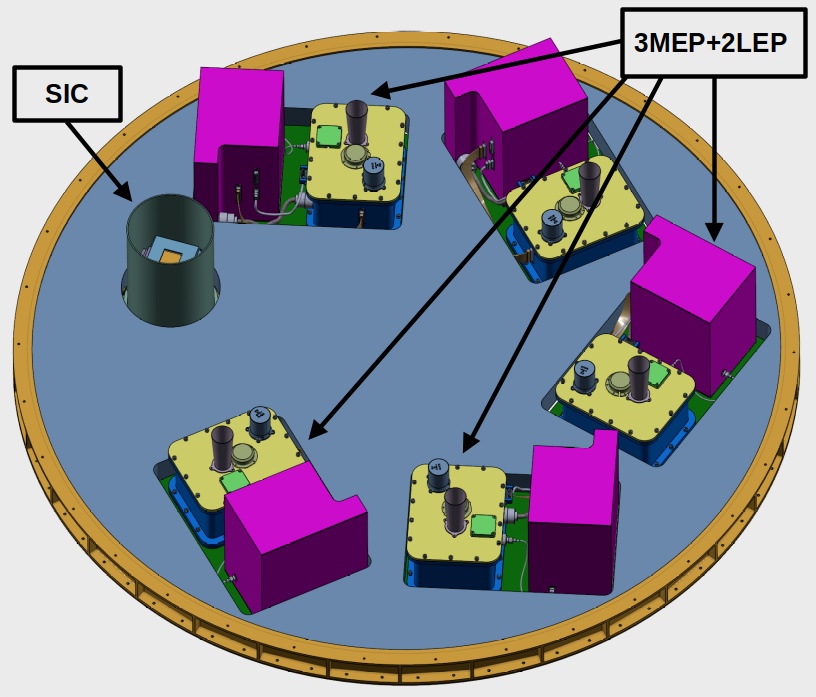}
\caption{The FPSS with attached the XRPs (2-LEPs and 3 MEPs) and the SIC at the correct position to be at the focus of the X-ray mirrors as displayed in Figure \ref{fig:FrontPanel}}
\label{fig:SIC_MEP_and_FPA_FFSS}
\end{subfigure}
\end{figure}

The \textbf{Focal Plane Assembly (FPA)} comprises the Focal Plane Support Structure (FPSS), the five X-ray Polarimeters (XRPs), the Spectral Imaging Camera (SIC), and the Instrument Control Unit (ICU). Their placement in the focal plane is shown in Figure \ref{fig:SIC_MEP_and_FPA_FFSS}.
The FPSS provides mechanical support and alignment reference for focal-plane instruments, radiators, and thermal hardware. Based on the XIPE (X-ray Imaging Polarimetry Explorer, ESA M4 proposal) Phase-A design, it consists of a 60 mm-thick composite sandwich panel with a Nomex honeycomb core and two 1 mm-thick CFRP skins, enclosed by a structural titanium ring that interfaces the FPSS with the metering structure.

To simplify the Assembly, Integration, and Verification (AIV) activities of both the payload and the spacecraft, the FPSS has been designed to allow independent integration of the XRPs, SIC, and ICU as soon as they become available after their functional testing and ground calibration campaigns. Dedicated mounting apertures, inserts, and alignment features enable each unit to be installed and verified independently, without requiring simultaneous integration of all payload elements. This is an outcome of XIPE phase A. 

\section{Mission Profile}

The \expo mission is based on pointed observations of known and newly discovered X-ray sources. The nominal observing plan can be interrupted at any time by a rapid Target of Opportunity (ToO) request autonomously generated by the WFI and ICU, which commands the spacecraft to repoint toward the transient source and, after completion of the observation, automatically resumes the original observing program. Rapid ToOs rely on the hydrazine propulsion system to achieve the required slew rate, whereas slower ToOs can be performed using only the reaction wheels, without propellant consumption.
The most demanding requirement for the Attitude Control System (ACS) is the capability to perform rapid autonomous repointing toward Gamma-Ray Bursts and magnetar flares with an average slew rate of $\sim1^\circ$ s$^{-1}$. Such maneuvers require torques well beyond the capability of reaction wheels and are therefore achieved using hydrazine thrusters. Once the target is acquired, reaction wheels provide fine pointing and enable the slow return to the nominal attitude. 
Preliminary analyses indicate that the required slew performance can be achieved with acceptable propellant consumption, compatible with the mission lifetime and the expected $\sim80$ rapid repointings per year. 
The ACS also includes star trackers, Sun sensors, inertial measurement units, GNSS and magnetometers (for LEO operations).

The baseline mission foresees a low-Earth equatorial orbit at an altitude of about 600 km. However, a Sun--Earth L1 orbit is also under assessment. Such an orbit would nearly double the number of sources accessible each year, simplify mission operations by eliminating Earth occultations, provide a more stable thermal environment, and avoid end-of-life disposal constraints. These advantages must be balanced against the need for a more capable telecommunications system to support the larger downlink distance and increased uplink power requirements.

\begin{table}[htbp!]
\caption{Baseline mission parameters of \expo. Mass and power budgets are upper limits based on the L1 configuration.}
\label{tab:MissionParameters}
\centering
\footnotesize
\renewcommand{\arraystretch}{1.1}
\begin{tabular}{ll}
\toprule
\textbf{Parameter} & \textbf{Value / property} \\
\midrule
Orbit & Baseline: 600 km equatorial (optional: L1) \\
Launcher & Ariane 62 \\
Ground Station & Baseline Malindi BSC (ASI) \\
Telemetry & X-band downlink/uplink \\
Data Rate & 60 Gbit day$^{-1}$ \\
Sun Aspect Angle & $60^\circ$--$120^\circ$ ($\sim$50\% sky accessibility) \\
Rapid ToO capability & 80 triggers yr$^{-1}$ (Swift-like, $\sim1^\circ$ s$^{-1}$ slew) \\
Mission Operations Centre (MOC) & ESOC, Darmstadt (Germany) \\
Science Operations Centre (SOC) & ESAC, Villafranca del Castillo (Spain) \\
Science Data Centre (SDC) & University of Geneva \\
Mirrors \& End-to-end Calibration Facility & PANTER X-ray Test Facility (MPE, Germany) \\
Mass budget (dry) & 2428 kg (including 20\% system margin) \\
Power budget & 2073 W (including 20\% system margin) \\
Launch & Baseline: 2041 \\
Mission lifetime & 5 years (goal: 7 years) \\
\bottomrule
\end{tabular}
\end{table}

\section{A summary of EXPO scientific goals}
An X-ray polarimetry mission combining rapid autonomous repointing with broadband sensitivity will revolutionize X-ray astrophysics, Time-Domain Astronomy, and Multi-Messenger Astronomy. The primary scientific goals of \expo\ include broadband polarimetry of the prompt emission of gamma-ray bursts (GRBs) down to 2~keV, probing the origin of the low-energy spectral break observed by \textit{Fermi} and \textit{Swift} and constraining the energy dissipation mechanisms operating in ultra-relativistic jets. In addition, polarimetry of the GRB afterglow, including the plateau and reflaring phases, will provide unique insight into the role of magnetic fields in powering the long-term evolution of these explosions.

Complementing the polarimetric measurements, the Spectroscopic Imaging Calorimeter (SIC) will perform high-quality spectroscopy of GRB afterglows up to the hard X-ray band, extending detailed spectroscopic studies into an energy range that has so far remained only sparsely explored.

Should intermediate flares occur in magnetars, \expo\ will enable a definitive test of quantum electrodynamics vacuum birefringence, owing to the large emitting area involved during these events. Simultaneously, broadband observations will investigate the connection between burst activity and the enigmatic hard X-ray power-law component.

Although requiring less stringent repointing times, time-resolved broadband polarimetry of transient black-hole X-ray binaries will reveal the origin of relativistic jets and clarify the interplay between the accretion disk, corona, and jet during spectral-state transitions.

The combination of broadband coverage and moderately rapid repointing will also enable detailed studies of X-ray pulsars as they evolve between sub-critical and super-critical accretion regimes, probing the distinct accretion geometries expected in these two states.

During short-lived blazar flares, time-resolved broadband X-ray polarimetry will discriminate between leptonic scenarios, characterized by rapid variability and inverse-Compton emission, and hadronic scenarios, dominated by slower variability and proton-synchrotron emission.

Several key science objectives do not require rapid repointing. Broadband polarimetry of X-ray reflection from black-hole and neutron-star binaries, molecular clouds surrounding the Galactic Centre, and active galactic nuclei—particularly Seyfert~2 and Compton-thick systems—will probe the geometry and physics of the reprocessing material. Polarimetric measurements above the fluorescent iron-line complex will provide a direct test of the contribution of returning radiation to the reflected emission from black-hole accretion disks. The extended energy coverage will also allow investigation of black-hole binaries exhibiting a polarization degree that rises steeply with energy, offering a powerful diagnostic of their coronal geometry and strong-gravity effects.

In accreting X-ray pulsars, polarimetry across cyclotron resonant scattering features (CRSFs) will probe radiative transfer in strongly magnetized plasmas, helping to explain the unexpectedly low polarization recently measured in several systems. Broadband observations of magnetars will firmly establish, on a statistical basis, the origin of their mysterious hard X-ray tails, extending the tantalizing but low-significance evidence obtained by \ixpe\ for a single source.

Finally, the angular resolution of \expo, approximately three times better than that of \ixpe\ in the common energy band and significantly improved over \textit{NuSTAR} in the overlapping hard X-ray range, will enable spatially resolved broadband polarimetry of extended sources. In supernova remnants, \expo\ will isolate regions with negligible thermal emission, mapping magnetic turbulence and particle acceleration with unprecedented detail. In pulsar wind nebulae, broadband imaging polarimetry will resolve the polarization properties of the inner nebula as the pulsar contribution becomes increasingly dominant at higher energies, providing new insight into particle acceleration and magnetic-field structure in these extreme environments.

\section{Conclusions}
\expo\ is a mission concept designed to overcome the narrow energy coverage and slow repointing of \ixpe while building on and extending its scientific legacy. 
\expo will combine broadband photoelectric polarimeters operating over 2--35 keV, a Spectral Imaging Camera for simultaneous spectroscopy, a Wide Field Instrument for continuous transient monitoring, and an autonomous repointing capability (SWIFT-like) reaching $\sim1^\circ$ s$^{-1}$.
This will deliver a four-times wider energy band, about three times larger effective area, and repointing times thousands of times faster than \ixpe.

These capabilities directly address several open questions in high-energy astrophysics: the origin of the hard X-ray tails of magnetars, the interplay between the inner flow and the relativistic jets in black-hole binaries, particle acceleration and turbulence in supernova remnants and pulsar-wind nebulae, X-ray reflection in AGNs including Compton-thick sources, and, uniquely, time-resolved polarimetry of GRB prompt and afterglow emission and the intermediate flares of magnetars. The payload relies on flight-proven or high-TRL technologies: GridPix detectors, gold-plated Ni-electroformed Wolter-I optics, and a coded-mask WFI derived from \textit{SVOM}/ECLAIRs, ensuring a low-risk, high-heritage development path towards a an ESA proposed launch around 2041.

By transforming X-ray polarimetry from a discovery-class capability into a mature observational tool, \expo\ will place Europe at the forefront of high-energy, time-domain, and multi-messenger astrophysics in the post-\ixpe era.

\acknowledgments % equivalent to \section*{ACKNOWLEDGMENTS}       
VEG acknowledges funding under NASA contract 80NSSC24K1403.
M.D., J.Pod., and J.S. acknowledge the support from the GACR project 26-22614S and the institutional support from RVO:67985815.
AV, JPou and ST acknowledge support from the Research Council of Finland grants 355672 and 372881 and Centre of Excellence in Neutron-Star Physics (grant 374064).
% References
\bibliography{References} % bibliography data in report.bib

@INPROCEEDINGS{Baracchini2026,
       author = {{Baracchini}, Elisabetta and {Dho}, Giorgio and {Fiorina}, Giorgio and {Soffitta}, Paolo and {Piacentini}, Stefano and {Costa}, Enrico and {Muleri}, Fabio and {Soffitta}, Paolo},
        title = "{Background Simulation for Future X-ray Polarimetry 1
Missions}",
    booktitle = {Space Telescopes and Instrumentation 2026: Ultraviolet to Gamma Ray, (to be published)},
         year = 2026,
       editor = {{Nikzad}, Shouleh  and {Kazuhiro}, Nakazawa and {Feroci}, Marco},
       series = {Society of Photo-Optical Instrumentation Engineers (SPIE) Conference Series},             
}

@ARTICLE{Dinsmore2024,
       author = {{Dinsmore}, Jack T. and {Romani}, Roger W.},
        title = "{A Catalog of Pulsar X-Ray Filaments}",
      journal = {\apj},
         year = 2024,
        month = nov,
       volume = {976},
       number = {1},
          eid = {4},
        pages = {4},
          doi = {10.3847/1538-4357/ad8344},
archivePrefix = {arXiv},
       eprint = {2410.01807},
 primaryClass = {astro-ph.HE},
       adsurl = {https://ui.adsabs.harvard.edu/abs/2024ApJ...976....4D}
}

@ARTICLE{Mikusincova2026,
       author = {{Miku{\v{s}}incov{\'a}}, Romana and {Veledina}, Alexandra and {Muleri}, Fabio and {Ciancarella}, Raul and {Zdziarski}, Andrzej and {Green}, David A. and {McCollough}, Michael and {Krawczynski}, Henric and {Steiner}, James F. and {Dov{\v{c}}iak}, Michal and {Ahlberg}, Varpu and {Bianchi}, Stefano and {Di Marco}, Alessandro and {Garc{\'\i}a}, Javier A. and {Ingram}, Adam and {Kaaret}, Philip and {Kallman}, Timothy and {Kun}, Hu and {La Monaca}, Fabio and {Lange}, Alexander and {Loktev}, Vladislav and {Mastroserio}, Guglielmo and {Matt}, Giorgio and {Emami}, Razieh and {Petrucci}, Pierre-Olivier and {Podgorn{\'y}}, Jakub and {Poutanen}, Juri and {Ratheesh}, Ajay and {Rodriguez}, Nicole and {Svoboda}, Ji{\v{r}}{\'\i} and {Tombesi}, Francesco and {Ursini}, Francesco and {Agudo}, Iv{\'a}n and {Antonelli}, Lucio A. and {Bachetti}, Matteo and {Baldini}, Luca and {Baumgartner}, Wayne H. and {Bellazzini}, Ronaldo and {Bongiorno}, Stephen D. and {Bonino}, Raffaella and {Brez}, Alessandro and {Bucciantini}, Niccol{\`o} and {Capitanio}, Fiamma and {Castellano}, Simone and {Cavazzuti}, Elisabetta and {Chen}, Chien-Ting and {Ciprini}, Stefano and {Costa}, Enrico and {De Rosa}, Alessandra and {Del Monte}, Ettore and {Di Gesu}, Laura and {Di Lalla}, Niccol{\`o} and {Donnarumma}, Immacolata and {Doroshenko}, Victor and {Ehlert}, Steven R. and {Enoto}, Teruaki and {Evangelista}, Yuri and {Fabiani}, Sergio and {Ferrazzoli}, Riccardo and {Gunji}, Shuichi and {Hayashida}, Kiyoshi and {Heyl}, Jeremy and {Iwakiri}, Wataru and {Jorstad}, Svetlana G. and {Karas}, Vladimir and {Kislat}, Fabian and {Kitaguchi}, Takao and {Kolodziejczak}, Jeffery J. and {Latronico}, Luca and {Liodakis}, Ioannis and {Maldera}, Simone and {Manfreda}, Alberto and {Marin}, Fr{\'e}d{\'e}ric and {Marinucci}, Andrea and {Marscher}, Alan P. and {Marshall}, Herman L. and {Massaro}, Francesco and {Mitsuishi}, Ikuyuki and {Mizuno}, Tsunefumi and {Negro}, Michela and {Yung Ng}, Chi- and {O'Dell}, Stephen L. and {Omodei}, Nicola and {Oppedisano}, Chiara and {Papitto}, Alessandro and {Pavlov}, George G. and {Peirson}, Abel L. and {Perri}, Matteo and {Pesce-Rollins}, Melissa and {Pilia}, Maura and {Possenti}, Andrea and {Puccetti}, Simonetta and {Ramsey}, Brian D. and {Rankin}, John and {Roberts}, Oliver J. and {Romani}, Roger W. and {Sgr{\`o}}, Carmelo and {Slane}, Patrick and {Soffitta}, Paolo and {Spandre}, Gloria and {Swartz}, Douglas A. and {Tamagawa}, Toru and {Tavecchio}, Fabrizio and {Taverna}, Roberto and {Tawara}, Yuzuru and {Tennant}, Allyn F. and {Thomas}, Nicholas E. and {Trois}, Alessio and {Tsygankov}, Sergey S. and {Turolla}, Roberto and {Vink}, Jacco and {Weisskopf}, Martin C. and {Wu}, Kinwah and {Xie}, Fei and {Zane}, Silvia},
        title = "{Super-Eddington Accretion Geometry: A Remarkable Stability of the Hidden Ultraluminous X-Ray Source Cygnus X-3}",
      journal = {\apj},
         year = 2026,
        month = jul,
       volume = {1005},
       number = {1},
          eid = {83},
        pages = {83},
          doi = {10.3847/1538-4357/ae7366},
archivePrefix = {arXiv},
       eprint = {2512.12879},
 primaryClass = {astro-ph.HE},
       adsurl = {https://ui.adsabs.harvard.edu/abs/2026ApJ..1005...83M}
}

@ARTICLE{Caiazzo2021,
       author = {{Caiazzo}, Ilaria and {Heyl}, Jeremy},
        title = "{Polarization of accreting X-ray pulsars. I. A new model}",
      journal = {\mnras},
         year = 2021,
        month = jan,
       volume = {501},
       number = {1},
        pages = {109-128},
          doi = {10.1093/mnras/staa3428},
archivePrefix = {arXiv},
       eprint = {2009.00631},
 primaryClass = {astro-ph.HE},
       adsurl = {https://ui.adsabs.harvard.edu/abs/2021MNRAS.501..109C}
}

@ARTICLE{Caiazzo2021b,
       author = {{Caiazzo}, Ilaria and {Heyl}, Jeremy},
        title = "{Polarization of accreting X-ray pulsars - II. Hercules X-1}",
      journal = {\mnras},
         year = 2021,
        month = jan,
       volume = {501},
       number = {1},
        pages = {129-136},
          doi = {10.1093/mnras/staa3429},
archivePrefix = {arXiv},
       eprint = {2009.00634},
 primaryClass = {astro-ph.HE},
       adsurl = {https://ui.adsabs.harvard.edu/abs/2021MNRAS.501..129C}
}

@ARTICLE{Gianolli2024,
       author = {{Gianolli}, V.~E. and {Bianchi}, S. and {Kammoun}, E. and {Gnarini}, A. and {Marinucci}, A. and {Ursini}, F. and {Parra}, M. and {Tortosa}, A. and {De Rosa}, A. and {Kim}, D.~E. and {Marin}, F. and {Matt}, G. and {Serafinelli}, R. and {Soffitta}, P. and {Tagliacozzo}, D. and {Di Gesu}, L. and {Done}, C. and {Marshall}, H.~L. and {Middei}, R. and {Mikusincova}, R. and {Petrucci}, P. -O. and {Ravi}, S. and {Svoboda}, J. and {Tombesi}, F.},
        title = "{A second view on the X-ray polarization of NGC 4151 with IXPE}",
      journal = {\aap},
         year = 2024,
        month = nov,
       volume = {691},
          eid = {A29},
        pages = {A29},
          doi = {10.1051/0004-6361/202451645},
archivePrefix = {arXiv},
       eprint = {2407.17243},
 primaryClass = {astro-ph.HE},
       adsurl = {https://ui.adsabs.harvard.edu/abs/2024A&A...691A..29G}
}

@ARTICLE{Gotz2006,
       author = {{G{\"o}tz}, D. and {Mereghetti}, S. and {Tiengo}, A. and {Esposito}, P.},
        title = "{Magnetars as persistent hard X-ray sources: INTEGRAL discovery of a hard tail in SGR 1900+14}",
      journal = {\aap},
         year = 2006,
        month = apr,
       volume = {449},
       number = {2},
        pages = {L31-L34},
          doi = {10.1051/0004-6361:20064870},
archivePrefix = {arXiv},
       eprint = {astro-ph/0602359},
 primaryClass = {astro-ph},
       adsurl = {https://ui.adsabs.harvard.edu/abs/2006A&A...449L..31G}
}

@ARTICLE{Halpern2014,
       author = {{Halpern}, J.~P. and {Tomsick}, J.~A. and {Gotthelf}, E.~V. and {Camilo}, F. and {Ng}, C.-Y. and {Bodaghee}, A. and {Rodriguez}, J. and {Chaty}, S. and {Rahoui}, F.},
        title = "{Discovery of X-Ray Pulsations from the INTEGRAL Source IGR J11014-6103}",
      journal = {\apjl},
         year = 2014,
        month = nov,
       volume = {795},
       number = {2},
          eid = {L27},
        pages = {L27},
          doi = {10.1088/2041-8205/795/2/L27},
archivePrefix = {arXiv},
       eprint = {1410.2332},
 primaryClass = {astro-ph.HE},
       adsurl = {https://ui.adsabs.harvard.edu/abs/2014ApJ...795L..27H}
}

@BOOK{Heitler1954,
   author = {{Heitler}, W.},
    title = "{Quantum theory of radiation}",
publisher = {International Series of Monographs on Physics, Oxford: Clarendon, 1954, 3rd ed.},
     year = 1954,
   adsurl = {http://adsabs.harvard.edu/abs/1954qtr..book.....H}
}

@INPROCEEDINGS{Khropost2025,
       author = {{Khropost}, Diana and {Riethm{\"u}ller}, Franziska and {D{\"o}hring}, Thorsten and {Flachs}, Dennis and {H{\"u}lag{\"u}}, Deniz and {Hertwig}, Andreas and {Cotroneo}, Vincenzo and {Gibertini}, Eugenio},
        title = "{Polydopamine: a bio-inspired polymer for x-ray mirror coatings and other technical applications}",
    booktitle = {EUV and X-ray Optics: Synergy between Laboratory and Space IX},
         year = 2025,
       editor = {{Hudec}, Ren{\'e} and {Pina}, Ladislav},
       series = {Society of Photo-Optical Instrumentation Engineers (SPIE) Conference Series},
       volume = {13531},
        month = jun,
          eid = {135310H},
        pages = {135310H},
          doi = {10.1117/12.3056298},
       adsurl = {https://ui.adsabs.harvard.edu/abs/2025SPIE13531E..0HK}
}

@ARTICLE{Krieger2013,
       author = {{Krieger}, C. and {Kaminski}, J. and {Desch}, K.},
        title = "{InGrid-based X-ray detector for low background searches}",
      journal = {Nuclear Instruments and Methods in Physics Research A},
         year = 2013,
        month = nov,
       volume = {729},
        pages = {905-909},
          doi = {10.1016/j.nima.2013.08.075},
       adsurl = {https://ui.adsabs.harvard.edu/abs/2013NIMPA.729..905K}
}

@ARTICLE{Muleri2026,
       author = {{Muleri}, Fabio and {Cesare}, Stefano and {Costa}, Enrico and {Cugno}, Walter and {Desch}, Klaus and {Di Marco}, Alessandro and {Fabiani}, Sergio and {Ferrazzoli}, Riccardo and {Gruber}, Markus and {Heuchel}, Daniel and {Imtiaz}, Saba and {Kaminski}, Jochen and {Kim}, Dawoon Edwin and {Lacerenza}, Alessandro and {Lefevre}, Carlo and {Manikantan}, Hemanth and {Plesanovs}, Vladislavs and {Rankin}, John and {Ratheesh}, Ajay and {Rubini}, Alda and {Soffitta}, Paolo},
        title = "{Instruments for Focal Plane X-Ray Polarimetry in the Next Decade}",
      journal = {Particles},
         year = 2026,
        month = mar,
       volume = {9},
       number = {2},
          eid = {30},
        pages = {30},
          doi = {10.3390/particles9020030},
archivePrefix = {arXiv},
       eprint = {2606.20337},
 primaryClass = {astro-ph.IM},
       adsurl = {https://ui.adsabs.harvard.edu/abs/2026Parti...9...30M}
}

@ARTICLE{Negro2023,
       author = {{Negro}, Michela and {Di Lalla}, Niccol{\`o} and {Omodei}, Nicola and {Veres}, P{\'e}ter and {Silvestri}, Stefano and {Manfreda}, Alberto and {Burns}, Eric and {Baldini}, Luca and {Costa}, Enrico and {Ehlert}, Steven R. and {Kennea}, Jamie A. and {Liodakis}, Ioannis and {Marshall}, Herman L. and {Mereghetti}, Sandro and {Middei}, Riccardo and {Muleri}, Fabio and {O'Dell}, Stephen L. and {Roberts}, Oliver J. and {Romani}, Roger W. and {Sgr{\'o}}, Carmelo and {Terashima}, Masanobu and {Tiengo}, Andrea and {Viscolo}, Domenico and {Di Marco}, Alessandro and {La Monaca}, Fabio and {Latronico}, Luca and {Matt}, Giorgio and {Perri}, Matteo and {Puccetti}, Simonetta and {Poutanen}, Juri and {Ratheesh}, Ajay and {Rogantini}, Daniele and {Slane}, Patrick and {Soffitta}, Paolo and {Lindfors}, Elina and {Nilsson}, Kari and {Kasikov}, Anni and {Marscher}, Alan P. and {Tavecchio}, Fabrizio and {Cibrario}, Nicol{\'o} and {Gunji}, Shuichi and {Malacaria}, Christian and {Paggi}, Alessandro and {Yang}, Yi-Jung and {Zane}, Silvia and {Weisskopf}, Martin C. and {Agudo}, Iv{\'a}n and {Antonelli}, Lucio A. and {Bachetti}, Matteo and {Baumgartner}, Wayne H. and {Bellazzini}, Ronaldo and {Bianchi}, Stefano and {Bongiorno}, Stephen D. and {Bonino}, Raffaella and {Brez}, Alessandro and {Bucciantini}, Niccol{\`o} and {Capitanio}, Fiamma and {Castellano}, Simone and {Cavazzuti}, Elisabetta and {Chen}, Chien-Ting and {Ciprini}, Stefano and {De Rosa}, Alessandra and {Del Monte}, Ettore and {Di Gesu}, Laura and {Donnarumma}, Immacolata and {Doroshenko}, Victor and {Dovc\v{c}iak}, Michal and {Enoto}, Teruaki and {Evangelista}, Yuri and {Fabiani}, Sergio and {Ferrazzoli}, Riccardo and {Garcia}, Javier A. and {Hayashida}, Kiyoshi and {Heyl}, Jeremy and {Iwakiri}, Wataru and {Jorstad}, Svetlana G. and {Kaaret}, Philip and {Karas}, Vladimir and {Kislat}, Fabian and {Kitaguchi}, Takao and {Kolodziejczak}, Jeffery J. and {Krawczynski}, Henric and {Maldera}, Simone and {Marin}, Fr{\'e}d{\'e}ric and {Marinucci}, Andrea and {Mitsuishi}, Ikuyuki and {Mizuno}, Tsunefumi and {Ng}, C.-Y. and {Oppedisano}, Chiara and {Papitto}, Alessandro and {Pavlov}, George G. and {Peirson}, Abel L. and {Pesce-Rollins}, Melissa and {Petrucci}, Pierre-Olivier and {Pilia}, Maura and {Possenti}, Andrea and {Ramsey}, Brian D. and {Rankin}, John and {Spandre}, Gloria and {Swartz}, Douglas A. and {Tamagawa}, Toru and {Taverna}, Roberto and {Tawara}, Yuzuru and {Tennant}, Allyn F. and {Thomas}, Nicholas E. and {Tombesi}, Francesco and {Trois}, Alessio and {Tsygankov}, Sergey S. and {Turolla}, Roberto and {Vink}, Jacco and {Wu}, Kinwah and {Xie}, Fei},
        title = "{The IXPE View of GRB 221009A}",
      journal = {\apjl},
         year = 2023,
        month = mar,
       volume = {946},
       number = {1},
          eid = {L21},
        pages = {L21},
          doi = {10.3847/2041-8213/acba17},
archivePrefix = {arXiv},
       eprint = {2301.01798},
 primaryClass = {astro-ph.HE},
       adsurl = {https://ui.adsabs.harvard.edu/abs/2023ApJ...946L..21N}
}

@ARTICLE{Ratheesh2024,
       author = {{Ratheesh}, Ajay and {Dov{\v{c}}iak}, Michal and {Krawczynski}, Henric and {Podgorn{\'y}}, Jakub and {Marra}, Lorenzo and {Veledina}, Alexandra and {Suleimanov}, Valery F. and {Rodriguez Cavero}, Nicole and {Steiner}, James F. and {Svoboda}, Ji{\v{r}}{\'\i} and {Marinucci}, Andrea and {Bianchi}, Stefano and {Negro}, Michela and {Matt}, Giorgio and {Tombesi}, Francesco and {Poutanen}, Juri and {Ingram}, Adam and {Taverna}, Roberto and {West}, Andrew and {Karas}, Vladimir and {Ursini}, Francesco and {Soffitta}, Paolo and {Capitanio}, Fiamma and {Viscolo}, Domenico and {Manfreda}, Alberto and {Muleri}, Fabio and {Parra}, Maxime and {Beheshtipour}, Banafsheh and {Chun}, Sohee and {Cibrario}, Nicol{\`o} and {Di Lalla}, Niccol{\`o} and {Fabiani}, Sergio and {Hu}, Kun and {Kaaret}, Philip and {Loktev}, Vladislav and {Miku{\v{s}}incov{\'a}}, Romana and {Mizuno}, Tsunefumi and {Omodei}, Nicola and {Petrucci}, Pierre-Olivier and {Puccetti}, Simonetta and {Rankin}, John and {Zane}, Silvia and {Zhang}, Sixuan and {Agudo}, Iv{\'a}n and {Antonelli}, Lucio A. and {Bachetti}, Matteo and {Baldini}, Luca and {Baumgartner}, Wayne H. and {Bellazzini}, Ronaldo and {Bongiorno}, Stephen D. and {Bonino}, Raffaella and {Brez}, Alessandro and {Bucciantini}, Niccol{\`o} and {Castellano}, Simone and {Cavazzuti}, Elisabetta and {Chen}, Chien-Ting and {Ciprini}, Stefano and {Costa}, Enrico and {De Rosa}, Alessandra and {Del Monte}, Ettore and {Di Gesu}, Laura and {Di Marco}, Alessandro and {Donnarumma}, Immacolata and {Doroshenko}, Victor and {Ehlert}, Steven R. and {Enoto}, Teruaki and {Evangelista}, Yuri and {Ferrazzoli}, Riccardo and {Garcia}, Javier A. and {Gunji}, Shuichi and {Hayashida}, Kiyoshi and {Heyl}, Jeremy and {Iwakiri}, Wataru and {Jorstad}, Svetlana G. and {Kislat}, Fabian and {Kitaguchi}, Takao and {Kolodziejczak}, Jeffery J. and {La Monaca}, Fabio and {Latronico}, Luca and {Liodakis}, Ioannis and {Maldera}, Simone and {Marin}, Fr{\'e}d{\'e}ric and {Marscher}, Alan P. and {Marshall}, Herman L. and {Massaro}, Francesco and {Mitsuishi}, Ikuyuki and {Ng}, Stephen C. -Y. and {O'Dell}, Stephen L. and {Oppedisano}, Chiara and {Papitto}, Alessandro and {Pavlov}, George G. and {Peirson}, Abel L. and {Perri}, Matteo and {Pesce-Rollins}, Melissa and {Pilia}, Maura and {Possenti}, Andrea and {Ramsey}, Brian D. and {Roberts}, Oliver J. and {Romani}, Roger W. and {Sgr{\`o}}, Carmelo and {Slane}, Patrick and {Spandre}, Gloria and {Swartz}, Douglas A. and {Tamagawa}, Toru and {Tavecchio}, Fabrizio and {Tawara}, Yuzuru and {Tennant}, Allyn F. and {Thomas}, Nicholas E. and {Trois}, Alessio and {Tsygankov}, Sergey S. and {Turolla}, Roberto and {Vink}, Jacco and {Weisskopf}, Martin C. and {Wu}, Kinwah and {Xie}, Fei},
        title = "{X-Ray Polarization of the Black Hole X-Ray Binary 4U 1630{\textendash}47 Challenges the Standard Thin Accretion Disk Scenario}",
      journal = {\apj},
         year = 2024,
        month = mar,
       volume = {964},
       number = {1},
          eid = {77},
        pages = {77},
          doi = {10.3847/1538-4357/ad226e},
archivePrefix = {arXiv},
       eprint = {2304.12752},
 primaryClass = {astro-ph.HE},
       adsurl = {https://ui.adsabs.harvard.edu/abs/2024ApJ...964...77R}
}

@INPROCEEDINGS{Ratheesh2024b,
       author = {{Ratheesh}, Ajay and {Soffitta}, Paolo and {Costa}, Enrico and {Del Monte}, Ettore and {Di Marco}, Alessandro and {Di Persio}, Giuseppe and {Desch}, Klaus and {De Angelis}, Nicolas and {Kim}, Dawoon E. and {Fabiani}, Sergio and {Ferrazzoli}, Riccardo and {Gruber}, Markus and {Kaminski}, Jochen and {Imtiaz}, Saba and {La Monaca}, Fabio and {Lefevre}, Carlo and {Manikantan}, Hemanth and {Morbidini}, Alfredo and {Mikusincova}, Romana and {Muleri}, Fabio and {Plesanovs}, Vladislavs and {Rankin}, John and {Rubini}, Alda},
        title = "{The legacy of IXPE: directions towards a new generation of 3D photo-electric x-ray polarimetry missions}",
    booktitle = {Space Telescopes and Instrumentation 2024: Ultraviolet to Gamma Ray},
         year = 2024,
       editor = {{den Herder}, Jan-Willem A. and {Nikzad}, Shouleh and {Nakazawa}, Kazuhiro},
       series = {Society of Photo-Optical Instrumentation Engineers (SPIE) Conference Series},
       volume = {13093},
        month = aug,
          eid = {1309386},
        pages = {1309386},
          doi = {10.1117/12.3020552},
       adsurl = {https://ui.adsabs.harvard.edu/abs/2024SPIE13093E..86R}
}

@INPROCEEDINGS{Manikantan2026,
       author = {{Manikantan}, Hemanth and {Lefevre}, Carlo and {Petrucci}, Lorenzo and {Soffitta}, Paolo and {Muleri}, Fabio and {Costa}, Enrico and {Rubini}, Alda and {Plesanovs}, Vladislavs and {Gruber}, Markus and {Kaminski}, Jochen and {Desch}, Klaus and {Di Marco}, Alessandro and {Fabiani}, Sergio and {Ferrazzoli}, Riccardo and {Imtiaz}, Saba and {Kim}, Dawoon and {Lacerenza}, Alessandro and {Rankin}, John and {Ratheesh}, Ajay},
        title = "{A Focal-Plane X-ray Polarimeter with Spectral and Timing Capabilities for Future X-ray Polarimetry Missions}",
    booktitle = {Space Telescopes and Instrumentation 2026: Ultraviolet to Gamma Ray, (to be published)},
         year = 2026,
       editor = {{Nikzad}, Shouleh  and {Kazuhiro}, Nakazawa and {Feroci}, Marco},
       series = {Society of Photo-Optical Instrumentation Engineers (SPIE) Conference Series},             
}

@ARTICLE{Veledina2023,
       author = {{Veledina}, Alexandra and {Muleri}, Fabio and {Poutanen}, Juri and {Podgorn{\'y}}, Jakub and {Dov{\v{c}}iak}, Michal and {Capitanio}, Fiamma and {Churazov}, Eugene and {De Rosa}, Alessandra and {Di Marco}, Alessandro and {Forsblom}, Sofia and {Kaaret}, Philip and {Krawczynski}, Henric and {La Monaca}, Fabio and {Loktev}, Vladislav and {Lutovinov}, Alexander A. and {Molkov}, Sergey V. and {Mushtukov}, Alexander A. and {Ratheesh}, Ajay and {Rodriguez Cavero}, Nicole and {Steiner}, James F. and {Sunyaev}, Rashid A. and {Tsygankov}, Sergey S. and {Zdziarski}, Andrzej A. and {Bianchi}, Stefano and {Bright}, Joe S. and {Bursov}, Nikolaj and {Costa}, Enrico and {Egron}, Elise and {Garcia}, Javier A. and {Green}, David A. and {Gurwell}, Mark and {Ingram}, Adam and {Kajava}, Jari J.~E. and {Kale}, Ruta and {Kraus}, Alex and {Malyshev}, Denys and {Marin}, Fr{\'e}d{\'e}ric and {Matt}, Giorgio and {McCollough}, Michael and {Mereminskiy}, Ilia A. and {Nizhelsky}, Nikolaj and {Piano}, Giovanni and {Pilia}, Maura and {Pittori}, Carlotta and {Rao}, Ramprasad and {Righini}, Simona and {Soffitta}, Paolo and {Shevchenko}, Anton and {Svoboda}, Jiri and {Tombesi}, Francesco and {Trushkin}, Sergei and {Tsybulev}, Peter and {Ursini}, Francesco and {Weisskopf}, Martin C. and {Wu}, Kinwah and {Agudo}, Iv{\'a}n and {Antonelli}, Lucio A. and {Bachetti}, Matteo and {Baldini}, Luca and {Baumgartner}, Wayne H. and {Bellazzini}, Ronaldo and {Bongiorno}, Stephen D. and {Bonino}, Raffaella and {Brez}, Alessandro and {Bucciantini}, Niccol{\`o} and {Castellano}, Simone and {Cavazzuti}, Elisabetta and {Chen}, Chien-Ting and {Ciprini}, Stefano and {Del Monte}, Ettore and {Di Gesu}, Laura and {Di Lalla}, Niccol{\`o} and {Donnarumma}, Immacolata and {Doroshenko}, Victor and {Ehlert}, Steven R. and {Enoto}, Teruaki and {Evangelista}, Yuri and {Fabiani}, Sergio and {Ferrazzoli}, Riccardo and {Gunji}, Shuichi and {Hayashida}, Kiyoshi and {Heyl}, Jeremy and {Iwakiri}, Wataru and {Jorstad}, Svetlana G. and {Karas}, Vladimir and {Kislat}, Fabian and {Kitaguchi}, Takao and {Kolodziejczak}, Jeffery J. and {Latronico}, Luca and {Liodakis}, Ioannis and {Maldera}, Simone and {Manfreda}, Alberto and {Marinucci}, Andrea and {Marscher}, Alan P. and {Marshall}, Herman L. and {Massaro}, Francesco and {Mitsuishi}, Ikuyuki and {Mizuno}, Tsunefumi and {Negro}, Michela and {Ng}, Chi-Yung and {O'Dell}, Stephen L. and {Omodei}, Nicola and {Oppedisano}, Chiara and {Papitto}, Alessandro and {Pavlov}, George G. and {Peirson}, Abel L. and {Perri}, Matteo and {Pesce-Rollins}, Melissa and {Petrucci}, Pierre-Olivier and {Possenti}, Andrea and {Puccetti}, Simonetta and {Ramsey}, Brian D. and {Rankin}, John and {Roberts}, Oliver and {Romani}, Roger W. and {Sgr{\`o}}, Carmelo and {Slane}, Patrick and {Spandre}, Gloria and {Swartz}, Doug and {Tamagawa}, Toru and {Tavecchio}, Fabrizio and {Taverna}, Roberto and {Tawara}, Yuzuru and {Tennant}, Allyn F. and {Thomas}, Nicholas E. and {Trois}, Alessio and {Turolla}, Roberto and {Vink}, Jacco and {Xie}, Fei and {Zane}, Silvia},
        title = "{Astronomical puzzle Cyg X-3 is a hidden Galactic ultraluminous X-ray source}",
      journal = {arXiv e-prints},
         year = 2023,
        month = mar,
          eid = {arXiv:2303.01174},
        pages = {arXiv:2303.01174},
          doi = {10.48550/arXiv.2303.01174},
archivePrefix = {arXiv},
       eprint = {2303.01174},
 primaryClass = {astro-ph.HE},
       adsurl = {https://ui.adsabs.harvard.edu/abs/2023arXiv230301174V}
}

@ARTICLE{Weisskopf2022,
       author = {{Weisskopf}, Martin C. and {Soffitta}, Paolo and {Baldini}, Luca and {Ramsey}, Brian D. and {O'Dell}, Stephen L. and {Romani}, Roger W. and {Matt}, Giorgio and {Deininger}, William D. and {Baumgartner}, Wayne H. and {Bellazzini}, Ronaldo and {Costa}, Enrico and {Kolodziejczak}, Jeffery J. and {Latronico}, Luca and {Marshall}, Herman L. and {Muleri}, Fabio and {Bongiorno}, Stephen D. and {Tennant}, Allyn and {Bucciantini}, Niccolo and {Dovciak}, Michal and {Marin}, Frederic and {Marscher}, Alan and {Poutanen}, Juri and {Slane}, Pat and {Turolla}, Roberto and {Kalinowski}, William and {Di Marco}, Alessandro and {Fabiani}, Sergio and {Minuti}, Massimo and {La Monaca}, Fabio and {Pinchera}, Michele and {Rankin}, John and {Sgro'}, Carmelo and {Trois}, Alessio and {Xie}, Fei and {Alexander}, Cheryl and {Allen}, D. Zachery and {Amici}, Fabrizio and {Andersen}, Jason and {Antonelli}, Angelo and {Antoniak}, Spencer and {Attin{\`a}}, Primo and {Barbanera}, Mattia and {Bachetti}, Matteo and {Baggett}, Randy M. and {Bladt}, Jeff and {Brez}, Alessandro and {Bonino}, Raffaella and {Boree}, Christopher and {Borotto}, Fabio and {Breeding}, Shawn and {Brienza}, Daniele and {Bygott}, H. Kyle and {Caporale}, Ciro and {Cardelli}, Claudia and {Carpentiero}, Rita and {Castellano}, Simone and {Castronuovo}, Marco and {Cavalli}, Luca and {Cavazzuti}, Elisabetta and {Ceccanti}, Marco and {Centrone}, Mauro and {Citraro}, Saverio and {D'Amico}, Fabio and {D'Alba}, Elisa and {Di Gesu}, Laura and {Del Monte}, Ettore and {Dietz}, Kurtis L. and {Di Lalla}, Niccolo' and {Persio}, Giuseppe Di and {Dolan}, David and {Donnarumma}, Immacolata and {Evangelista}, Yuri and {Ferrant}, Kevin and {Ferrazzoli}, Riccardo and {Ferrie}, MacKenzie and {Footdale}, Joseph and {Forsyth}, Brent and {Foster}, Michelle and {Garelick}, Benjamin and {Gunji}, Shuichi and {Gurnee}, Eli and {Head}, Michael and {Hibbard}, Grant and {Johnson}, Samantha and {Kelly}, Erik and {Kilaru}, Kiranmayee and {Lefevre}, Carlo and {Roy}, Shelley Le and {Loffredo}, Pasqualino and {Lorenzi}, Paolo and {Lucchesi}, Leonardo and {Maddox}, Tyler and {Magazzu}, Guido and {Maldera}, Simone and {Manfreda}, Alberto and {Mangraviti}, Elio and {Marengo}, Marco and {Marrocchesi}, Alessandra and {Massaro}, Francesco and {Mauger}, David and {McCracken}, Jeffrey and {McEachen}, Michael and {Mize}, Rondal and {Mereu}, Paolo and {Mitchell}, Scott and {Mitsuishi}, Ikuyuki and {Morbidini}, Alfredo and {Mosti}, Federico and {Nasimi}, Hikmat and {Negri}, Barbara and {Negro}, Michela and {Nguyen}, Toan and {Nitschke}, Isaac and {Nuti}, Alessio and {Onizuka}, Mitch and {Oppedisano}, Chiara and {Orsini}, Leonardo and {Osborne}, Darren and {Pacheco}, Richard and {Paggi}, Alessandro and {Painter}, Will and {Pavelitz}, Steven D. and {Pentz}, Christina and {Piazzolla}, Raffaele and {Perri}, Matteo and {Pesce-Rollins}, Melissa and {Peterson}, Colin and {Pilia}, Maura and {Profeti}, Alessandro and {Puccetti}, Simonetta and {Ranganathan}, Jaganathan and {Ratheesh}, Ajay and {Reedy}, Lee and {Root}, Noah and {Rubini}, Alda and {Ruswick}, Stephanie and {Sanchez}, Javier and {Sarra}, Paolo and {Santoli}, Francesco and {Scalise}, Emanuele and {Sciortino}, Andrea and {Schroeder}, Christopher and {Seek}, Tim and {Sosdian}, Kalie and {Spandre}, Gloria and {Speegle}, Chet O. and {Tamagawa}, Toru and {Tardiola}, Marcello and {Tobia}, Antonino and {Thomas}, Nicholas E. and {Valerie}, Robert and {Vimercati}, Marco and {Walden}, Amy L. and {Weddendorf}, Bruce and {Wedmore}, Jeffrey and {Welch}, David and {Zanetti}, Davide and {Zanetti}, Francesco},
        title = "{The Imaging X-Ray Polarimetry Explorer (IXPE): Pre-Launch}",
      journal = {Journal of Astronomical Telescopes, Instruments, and Systems},
         year = 2022,
        month = apr,
       volume = {8},
       number = {2},
          eid = {026002},
        pages = {026002},
          doi = {10.1117/1.JATIS.8.2.026002},
archivePrefix = {arXiv},
       eprint = {2112.01269},
 primaryClass = {astro-ph.IM},
       adsurl = {https://ui.adsabs.harvard.edu/abs/2022JATIS...8b6002W}
}

@ARTICLE{Veledina2024,
       author = {{Veledina}, Alexandra and {Muleri}, Fabio and {Poutanen}, Juri and {Podgorn{\'y}}, Jakub and {Dov{\v{c}}iak}, Michal and {Capitanio}, Fiamma and {Churazov}, Eugene and {De Rosa}, Alessandra and {Di Marco}, Alessandro and {Forsblom}, Sofia V. and {Kaaret}, Philip and {Krawczynski}, Henric and {La Monaca}, Fabio and {Loktev}, Vladislav and {Lutovinov}, Alexander A. and {Molkov}, Sergey V. and {Mushtukov}, Alexander A. and {Ratheesh}, Ajay and {Rodriguez Cavero}, Nicole and {Steiner}, James F. and {Sunyaev}, Rashid A. and {Tsygankov}, Sergey S. and {Weisskopf}, Martin C. and {Zdziarski}, Andrzej A. and {Bianchi}, Stefano and {Bright}, Joe S. and {Bursov}, Nikolaj and {Costa}, Enrico and {Egron}, Elise and {Garcia}, Javier A. and {Green}, David A. and {Gurwell}, Mark and {Ingram}, Adam and {Kajava}, Jari J.~E. and {Kale}, Ruta and {Kraus}, Alex and {Malyshev}, Denys and {Marin}, Fr{\'e}d{\'e}ric and {Matt}, Giorgio and {McCollough}, Michael and {Mereminskiy}, Ilya A. and {Nizhelsky}, Nikolaj and {Piano}, Giovanni and {Pilia}, Maura and {Pittori}, Carlotta and {Rao}, Ramprasad and {Righini}, Simona and {Soffitta}, Paolo and {Shevchenko}, Anton and {Svoboda}, Jiri and {Tombesi}, Francesco and {Trushkin}, Sergei A. and {Tsybulev}, Peter and {Ursini}, Francesco and {Wu}, Kinwah and {Agudo}, Iv{\'a}n and {Antonelli}, Lucio A. and {Bachetti}, Matteo and {Baldini}, Luca and {Baumgartner}, Wayne H. and {Bellazzini}, Ronaldo and {Bongiorno}, Stephen D. and {Bonino}, Raffaella and {Brez}, Alessandro and {Bucciantini}, Niccol{\`o} and {Castellano}, Simone and {Cavazzuti}, Elisabetta and {Chen}, Chien-Ting and {Ciprini}, Stefano and {Del Monte}, Ettore and {Di Gesu}, Laura and {Di Lalla}, Niccol{\`o} and {Donnarumma}, Immacolata and {Doroshenko}, Victor and {Ehlert}, Steven R. and {Enoto}, Teruaki and {Evangelista}, Yuri and {Fabiani}, Sergio and {Ferrazzoli}, Riccardo and {Gunji}, Shuichi and {Hayashida}, Kiyoshi and {Heyl}, Jeremy and {Iwakiri}, Wataru and {Jorstad}, Svetlana G. and {Karas}, Vladimir and {Kislat}, Fabian and {Kitaguchi}, Takao and {Kolodziejczak}, Jeffery J. and {Latronico}, Luca and {Liodakis}, Ioannis and {Maldera}, Simone and {Manfreda}, Alberto and {Marinucci}, Andrea and {Marscher}, Alan P. and {Marshall}, Herman L. and {Massaro}, Francesco and {Mitsuishi}, Ikuyuki and {Mizuno}, Tsunefumi and {Negro}, Michela and {Ng}, Chi-Yung and {O'Dell}, Stephen L. and {Omodei}, Nicola and {Oppedisano}, Chiara and {Papitto}, Alessandro and {Pavlov}, George G. and {Peirson}, Abel L. and {Perri}, Matteo and {Pesce-Rollins}, Melissa and {Petrucci}, Pierre-Olivier and {Possenti}, Andrea and {Puccetti}, Simonetta and {Ramsey}, Brian D. and {Rankin}, John and {Roberts}, Oliver and {Romani}, Roger W. and {Sgr{\`o}}, Carmelo and {Slane}, Patrick and {Spandre}, Gloria and {Swartz}, Doug and {Tamagawa}, Toru and {Tavecchio}, Fabrizio and {Taverna}, Roberto and {Tawara}, Yuzuru and {Tennant}, Allyn F. and {Thomas}, Nicholas E. and {Trois}, Alessio and {Turolla}, Roberto and {Vink}, Jacco and {Xie}, Fei and {Zane}, Silvia},
        title = "{Cygnus X-3 revealed as a Galactic ultraluminous X-ray source by IXPE}",
      journal = {Nature Astronomy},
         year = 2024,
        month = aug,
       volume = {8},
        pages = {1031-1046},
          doi = {10.1038/s41550-024-02294-9},
archivePrefix = {arXiv},
       eprint = {2303.01174},
 primaryClass = {astro-ph.HE},
       adsurl = {https://ui.adsabs.harvard.edu/abs/2024NatAs...8.1031V}
}

@ARTICLE{Veledina2024b,
       author = {{Veledina}, Alexandra and {Poutanen}, Juri and {Bocharova}, Anastasiia and {Di Marco}, Alessandro and {Forsblom}, Sofia V. and {La Monaca}, Fabio and {Podgorn{\'y}}, Jakub and {Tsygankov}, Sergey S. and {Zdziarski}, Andrzej A. and {Ahlberg}, Varpu and {Green}, David A. and {Muleri}, Fabio and {Rhodes}, Lauren and {Bianchi}, Stefano and {Costa}, Enrico and {Dov{\v{c}}iak}, Michal and {Loktev}, Vladislav and {McCollough}, Michael and {Soffitta}, Paolo and {Sunyaev}, Rashid},
        title = "{Ultrasoft state of microquasar Cygnus X-3: X-ray polarimetry reveals the geometry of the astronomical puzzle}",
      journal = {\aap},
         year = 2024,
        month = aug,
       volume = {688},
          eid = {L27},
        pages = {L27},
          doi = {10.1051/0004-6361/202451356},
archivePrefix = {arXiv},
       eprint = {2407.02655},
 primaryClass = {astro-ph.HE},
       adsurl = {https://ui.adsabs.harvard.edu/abs/2024A&A...688L..27V}
}

@ARTICLE{Atteia2022,
       author = {{Atteia}, J.-L. and {Cordier}, B. and {Wei}, J.},
        title = "{The SVOM mission}",
      journal = {International Journal of Modern Physics D},
         year = 2022,
        month = apr,
       volume = {31},
       number = {5},
          eid = {2230008},
        pages = {2230008},
          doi = {10.1142/S0218271822300087},
archivePrefix = {arXiv},
       eprint = {2203.10962},
 primaryClass = {astro-ph.IM},
       adsurl = {https://ui.adsabs.harvard.edu/abs/2022IJMPD..3130008A}
}

@INPROCEEDINGS{Cotroneo2022,
       author = {{Cotroneo}, Vincenzo and {Rivolta}, Giacomo and {Bavdaz}, Marcos and {Bruni}, Ricardo and {Civitani}, Marta M. and {D{\"o}hring}, Thorsten and {Ferreira}, Ivo and {Gibertini}, Eugenio and {Giglia}, Angelo and {Gollwitzer}, Christian and {Iovenitti}, Simone and {Krumney}, Michael and {Magagnin}, Luca and {Mahne}, Nicola and {Nannarone}, Stefano and {Pareschi}, Giovanni and {Romaine}, Suzanne and {Sethares}, Leandra and {Shortt}, Brian and {Skroblin}, Dieter and {Sironi}, Giorgia and {Spiga}, Daniele and {Tagliaferri}, Gianpiero and {Valsecchi}, Giuseppe},
        title = "{Dopamine dip-liquid overcoatings for soft x-ray reflectivity enhancement}",
    booktitle = {Space Telescopes and Instrumentation 2022: Ultraviolet to Gamma Ray},
         year = 2022,
       editor = {{den Herder}, Jan-Willem A. and {Nikzad}, Shouleh and {Nakazawa}, Kazuhiro},
       series = {Society of Photo-Optical Instrumentation Engineers (SPIE) Conference Series},
       volume = {12181},
        month = aug,
          eid = {1218117},
        pages = {1218117},
          doi = {10.1117/12.2630212},
       adsurl = {https://ui.adsabs.harvard.edu/abs/2022SPIE12181E..17C}
}

@ARTICLE{vanderGraaf2007,
       author = {{van der Graaf}, Harry},
        title = "{GridPix: An integrated readout system for gaseous detectors with a pixel chip as anode}",
      journal = {Nuclear Instruments and Methods in Physics Research A},
         year = 2007,
        month = oct,
       volume = {580},
        pages = {1023-1026},
          doi = {10.1016/j.nima.2007.06.096},
       adsurl = {https://ui.adsabs.harvard.edu/abs/2007NIMPA.580.1023V}
}

@ARTICLE{Kaminski2012,
       author = {{Kaminski}, J. and {Brezina}, C. and {Desch}, K. and {Killenberg}, M. and {Krautscheid}, T. and {Krieger}, C. and {M{\"u}ller}, F. and {Schultens}, M.},
        title = "{Gaseous detectors with micropattern gas amplification stages and CMOS pixel chip readout}",
      journal = {Journal of Instrumentation},
         year = 2012,
        month = feb,
       volume = {7},
       number = {2},
        pages = {C02035},
          doi = {10.1088/1748-0221/7/02/C02035},
       adsurl = {https://ui.adsabs.harvard.edu/abs/2012JInst...7C2035K}
}

@ARTICLE{Poikela2014,
       author = {{Poikela}, T. and {Plosila}, J. and {Westerlund}, T. and {Campbell}, M. and {De Gaspari}, M. and {Llopart}, X. and {Gromov}, V. and {Kluit}, R. and {van Beuzekom}, M. and {Zappon}, F. and {Zivkovic}, V. and {Brezina}, C. and {Desch}, K. and {Fu}, Y. and {Kruth}, A.},
        title = "{Timepix3: a 65K channel hybrid pixel readout chip with simultaneous ToA/ToT and sparse readout}",
      journal = {Journal of Instrumentation},
         year = 2014,
        month = may,
       volume = {9},
       number = {5},
          eid = {C05013},
        pages = {C05013},
          doi = {10.1088/1748-0221/9/05/C05013},
       adsurl = {https://ui.adsabs.harvard.edu/abs/2014JInst...9C5013P}
}

@ARTICLE{2003Granot,
       author = {{Granot}, Jonathan and {K{\"o}nigl}, Arieh},
        title = "{Linear Polarization in Gamma-Ray Bursts: The Case for an Ordered Magnetic Field}",
      journal = {\apjl},
         year = 2003,
        month = sep,
       volume = {594},
       number = {2},
        pages = {L83-L87},
          doi = {10.1086/378733},
archivePrefix = {arXiv},
       eprint = {astro-ph/0304286},
 primaryClass = {astro-ph},
       adsurl = {https://ui.adsabs.harvard.edu/abs/2003ApJ...594L..83G}
}

@ARTICLE{2025extp,
       author = {{Zhang}, Shuang-Nan and {Santangelo}, Andrea and {Xu}, Yupeng and {Feng}, Hua and {Lu}, Fangjun and {Chen}, Yong and {Ge}, Mingyu and {Nandra}, Kirpal and {Wu}, Xin and {Feroci}, Marco and {Hernanz}, Margarita and {Liu}, Congzhan and {He}, Huilin and {Wang}, Yusa and {Jiang}, Weichun and {Cui}, Weiwei and {Yang}, Yanji and {Wang}, Juan and {Li}, Wei and {Li}, Hong and {Du}, Yuanyuan and {Liu}, Xiaohua and {Meng}, Bin and {Wen}, Xiangyang and {Zhang}, Aimei and {Ma}, Jia and {Li}, Maoshun and {Li}, Gang and {Qi}, Liqiang and {Sun}, Jianchao and {Luo}, Tao and {Liu}, Hongwei and {Liu}, Xiaojing and {Zhang}, Fan and {Luo}, Laidan and {Zhu}, Yuxuan and {Zhao}, Zijian and {Sun}, Liang and {Yang}, Xiongtao and {Wu}, Qiong and {Jiang}, Jiechen and {Shi}, Haoli and {Liu}, Jiangtao and {Xu}, Yanbing and {Yang}, Sheng and {Zhang}, Laiyu and {Han}, Dawei and {Gao}, Na and {Huo}, Jia and {Zhang}, Ziliang and {Wang}, Hao and {Zhao}, Xiaofan and {Wang}, Shuo and {Li}, Zhenjie and {Bao}, Ziyu and {Liu}, Yaoguang and {Wang}, Ke and {Wang}, Na and {Wang}, Bo and {Wang}, Langping and {Wang}, Dianlong and {Ding}, Fei and {Sheng}, Lizhi and {Qiang}, Pengfei and {Yan}, Yongqing and {Liu}, Yongan and {Wu}, Zhenyu and {Liu}, Yichen and {Chen}, Hao and {Zhang}, Yacong and {Liu}, Hongbang and {Altmann}, Alexander and {Bechteler}, Thomas and {Burwitz}, Vadim and {Fiorini}, Carlo and {Friedrich}, Peter and {Meidinger}, Norbert and {Strecker}, Rafael and {Baldini}, Luca and {Bellazzini}, Ronaldo and {Bonino}, Raffaella and {Frass{\`a}}, Andrea and {Latronico}, Luca and {Maldera}, Simone and {Manfreda}, Alberto and {Minuti}, Massimo and {Pesce-Rollins}, Melissa and {Sgr{\`o}}, Carmelo and {Tugliani}, Stefano and {Pareschi}, Giovanni and {Basso}, Stefano and {Sironi}, Giorgia and {Spiga}, Daniele and {Tagliaferri}, Gianpiero and {Tykhonov}, Andrii and {Paltani}, St{\`e}phane and {Bozzo}, Enrico and {Tenzer}, Christoph and {Bayer}, J{\"o}rg and {Tuo}, Youli and {Liu}, Honghui and {Zhang}, Yonghe and {Cai}, Zhiming and {Liu}, Huaqiu and {Chen}, Wen and {Wang}, Chunhong and {He}, Tao and {Chen}, Yehai and {Qiu}, Chengbo and {Zhang}, Ye and {Feng}, Jianchao and {Zhu}, Xiaofei and {Zhou}, Heng and {Zheng}, Shijie and {Song}, Liming and {Wang}, Jinzhou and {Jia}, Shumei and {Jiang}, Zewen and {Li}, Xiaobo and {Zhao}, Haisheng and {Guan}, Ju and {Zhang}, Juan and {Li}, Chengkui and {Huang}, Yue and {Liao}, Jinyuan and {You}, Yuan and {Zhang}, Hongmei and {Wang}, Wenshuai and {Wang}, Shuang and {Ou}, Ge and {Hu}, Hao and {Shi}, Jingyan and {Cui}, Tao and {Jiang}, Xiaowei and {Cheng}, Yaodong and {Li}, Haibo and {Xu}, Yanjun and {Zane}, Silvia and {Bambi}, Cosimo and {Bu}, Qingcui and {Dall'Osso}, Simone and {Rosa}, Alessandra De and {Gou}, Lijun and {Guillot}, Sebastien and {Ji}, Long and {Li}, Ang and {Mao}, Jirong and {Patruno}, Alessandro and {Stratta}, Giulia and {Taverna}, Roberto and {Tsygankov}, Sergey and {Uttley}, Phil and {Watts}, Anna L. and {Wu}, Xuefeng and {Xu}, Renxin and {Yi}, Shuxu and {Zhang}, Guobao and {Zhang}, Liang and {Zhao}, Wen and {Zhou}, Ping},
        title = "{The enhanced X-ray Timing and Polarimetry mission{\textemdash}eXTP for launch in 2030}",
      journal = {Science China Physics, Mechanics, and Astronomy},
         year = 2025,
        month = sep,
       volume = {68},
       number = {11},
          eid = {119502},
        pages = {119502},
          doi = {10.1007/s11433-025-2786-6},
archivePrefix = {arXiv},
       eprint = {2506.08101},
 primaryClass = {astro-ph.HE},
       adsurl = {https://ui.adsabs.harvard.edu/abs/2025SCPMA..6819502Z}
}
\bibliographystyle{spiebib} % makes bibtex use spiebib.bst

\end{document}